\documentclass[journal]{IEEEtran}

\ifCLASSINFOpdf
\else
   \usepackage[dvips]{graphicx}
\fi
\usepackage{url}

\usepackage{graphicx}
\usepackage{amsfonts}
\usepackage{amsmath}
\usepackage{hyperref}
\usepackage{multirow}
\usepackage[dvipsnames]{xcolor}
\usepackage{colortbl}
\usepackage{array, makecell}
\usepackage{stmaryrd}
\definecolor{lightgray}{rgb}{0.9, 0.9, 0.9}

\definecolor{mycolor}{rgb}{0.12156862745098039, 0.4666666666666667, 0.7058823529411765} 

\usepackage{amsthm}
\newtheorem{lemma}{Lemma}
\newtheorem{definition}{Definition}
\newtheorem{proposition}{Proposition}

\usepackage{stackengine} 

\usepackage{tikz} 
\usetikzlibrary{shapes, shadows, arrows}
\usetikzlibrary{positioning}
\tikzset{mynode/.style={align=center,text width=3cm}}

\begin{document}

\title{DCT2net: an interpretable shallow CNN for image denoising}

\author{Sébastien Herbreteau and Charles Kervrann
\thanks{Sébastien Herbreteau and Charles Kervrann are with Inria Rennes - Bretagne Atlantique and UMR144-CNRS Institut Curie PSL, Paris, France (e-mail: sebastien.herbreteau@inria.fr and charles.kervrann@inria.fr)}}

\markboth{Journal of \LaTeX\ Class Files, Vol. 14, No. 8, August 2015}
{Shell \MakeLowercase{\textit{et al.}}: Bare Demo of IEEEtran.cls for IEEE Journals}
\maketitle

\begin{abstract}
This work tackles the issue of noise removal from images, focusing on the well-known DCT image denoising algorithm. The latter, stemming from signal processing, has been well studied over the years. Though very simple, it is still used in crucial parts of state-of-the-art "traditional" denoising algorithms such as BM3D. Since a few years however, deep convolutional neural networks (CNN) have outperformed their traditional counterparts, making signal processing methods less attractive. In this paper, we demonstrate that a DCT denoiser can be seen as a shallow CNN and thereby its original linear transform can be tuned through gradient descent in a supervised manner, improving considerably its performance. This gives birth to a fully interpretable CNN called DCT2net. To deal with remaining artifacts induced by DCT2net, an original hybrid solution between DCT and DCT2net is proposed combining the best that these two methods can offer; DCT2net is selected to process non-stationary image patches while DCT is optimal for piecewise smooth patches. Experiments on artificially noisy images demonstrate that two-layer DCT2net provides comparable results to BM3D and is as fast as DnCNN algorithm composed of more than a dozen of layers.
\end{abstract}


\IEEEpeerreviewmaketitle

\section{Introduction}

\IEEEPARstart{I}{mage} denoising is one of the most widely explored problems in computational imaging. In its most studied formulation, an image $\boldsymbol{x}$ is assumed to be corrupted by additive white  Gaussian noise (AWGN), $\boldsymbol{\varepsilon}$, with variance $\sigma^2$. The observed noisy image $\boldsymbol{y} = \boldsymbol{x} + \boldsymbol{\varepsilon}$ has then to be processed to recover the original signal $\boldsymbol{x}$ while removing the noise component $\boldsymbol{\varepsilon}$. 

Over the years, a rich variety of methods have been proposed to deal with this issue with inspiration coming from multiple fields. Frequency-based methods, aiming at decomposing the signal in a DCT or wavelet basis and then shrinking some transform coefficients, were considered quite early \cite{DCT, wavelet}. This strategy, coming from compression algorithms \cite{JPEG}, has the advantage to be both simple and fast but also the drawback of suppressing fine details in the image. An improvement of such methods consists in considering an overcomplete dictionary instead, under an assumption of sparse representation: each patch of an image is supposed to be sufficiently represented by a few vectors of an overcomplete basis \cite{ksvd}. In the meantime, the N(on)L(ocal)-means algorithm \cite{nlmeans} opened the door to a new category of denoising algorithms exploiting the self-similarity assumption. Indeed, it was observed that, within the same image, similar patches are repeated  in the whole image \cite{irani}. By gathering together the most similar patches and denoising them all at once, considerable gains in performance can be obtained \cite{BM3D, LSSC, nlbayes, WNNM, PEWA, OWF, optimalS, localA}. Most of those methods achieved state-of-the-art results until recently. 

In the last five years, the development of deep learning have revolutionized computer vision, through significant accuracy improvements, denoising task being no exception. A lot of convolutional neural networks have been proposed \cite{dncnn, ffdnet, red30, tnrd, nlrn, n3net, mlp} and they all outperformed the traditional algorithms via image training sets. Though fast and efficient, they all suffer from their lack of interpretability. Acting as "black boxes", it can be very challenging to thoroughly understand how they produce a result, which can be prohibitive for critical applications such as medical imaging.

Our work contributes to the recent trend, which builds on traditional algorithms and revisits them with a dose of deep learning, while keeping the original intuition \cite{deepKSVD, BM3Dnet}. We focus specifically on the DCT denoiser \cite{DCT} and show that it can be seen as a shallow CNN with weights corresponding to the DCT projection kernel and a hard shrinkage function as activation function. By training this particular CNN given external dataset, we can refine the resulting transform and boost its performance. As the so-called DCT2net  inherently may create unpleasant artifacts in flat regions of the image, we apply a two-class classification procedure based on the Canny egde detector \cite{canny} applied to the image denoised with the original DCT denoiser. The classification produces a binary map that separates homogeneous regions from textured regions and contours. DCT2net is then applied to the set of pixels with more complex geometries, while DCT is applied to stationary patches (i.e., with no significant spatial gradients). Surprisingly, this strategy does not alter performances in terms of Peak Signal-to-Noise Ratio (PSNR) while visually improving the results.

The remainder of the paper is organized as follows. In Section \ref{section2}, we present the principle of DCT denoiser and the properties of DCT2net which has the advantage to be invariant to the level of noise in image unlike other CNN-based denoisers \cite{dncnn, red30, tnrd}. In Section \ref{section3}, we interpret the "pseudo" basis learned by DCT2net and show that patch denoising and aggregation are jointly performed unlike traditional patch-based denoisers. In Section \ref{section4}, we analyze the behavior of DCT2net which is further mixed with the usual DCT denoiser to reduce unpleasant visual artifacts. In Section \ref{section5}, experimental results on datasets demonstrate that DCT2net is very fast as DnCNN, improves significantly the DCT results and is comparable to BM3D in terms of performance while remaining very simple and interpretable as the DCT denoiser.

\section{From popular DCT denoising to DCT2net} \label{section2}

In what follows, the vector representation of an image is adopted. A noisy 2D image $\boldsymbol{y}$ composed of $n$ pixels is formally represented by a vector of $\mathbb{R}^n$. 

\subsection{Traditional DCT denoiser}

In its most mature formulation for image denoising \cite{DCT}, the DCT denoiser proceeds on small overlapping patches accross the image. Each patch of size $p \times p$ is denoised independently, so that each pixel is in fact denoised $p^2$ times. For each pixel, the final denoised value is then obtained by averaging those $p^2$ estimators. Typical values for $p$ are powers of $2$ (generally $8$ or $16$) for practical reasons in the computation of the fast discrete cosine transform. However, we focus here on the case where $p$ is an odd number, without loss of generality.

\begin{figure}[t]
\centerline{\includegraphics[width=\columnwidth]{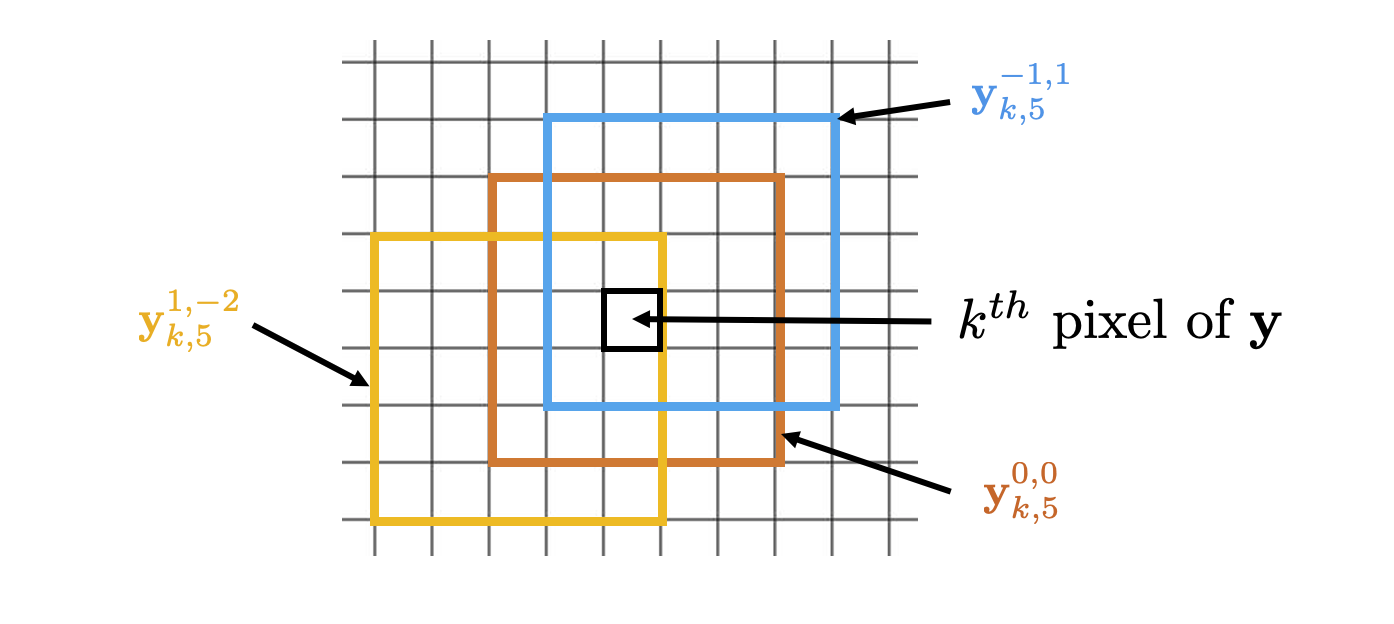}}
\caption{Some examples of the notation $\boldsymbol{y}_{k, p}^{i,j}$.}
 \label{notation}
\end{figure}

For the sake of representation, we denote by $\boldsymbol{y}_{k, p}^{i,j}$ the $p \times p$ patch, in the vector form, for which the central pixel, is located $j$ pixels at the right of the $k^{th}$ pixel of $\boldsymbol{y}$ and $i$ pixels beneath it (see Fig. \ref{notation}). Note that $i$ and $j$ can be negative numbers. Let us denote $q = \lfloor \frac{p}{2} \rfloor$ (i.e. closest integer less than or equal to $p/2$). The DCT denoiser denoted $F$ can then be expressed as: 
\begin{equation}
    F(\boldsymbol{y})_k = \frac{1}{p^2} \sum_{i = -q}^{q} \sum_{j = -q}^{q} [\boldsymbol{P} \varphi_{\lambda} (\boldsymbol{P}^{-1} \boldsymbol{y}_{k, p}^{i,j})]_{s(i, j)}
\label{dctdenoiser}
\end{equation}

\noindent where $\boldsymbol{P}$ is a matrix of size $p^2 \times p^2$, $\varphi_{\lambda}$ is the hard shrinkage function $\varphi_{\lambda}(x) = x \times \boldsymbol{1}_{\mathbb{R}\backslash[-\lambda, \lambda]}(x)$, and $s(i, j) = (q-i)p + q-j+1$ . According to \cite{DCT}, the most appropriate choice for $\lambda$ is $3 \sigma$.

By definition of DCT, the mathematical expression of the coefficient of the matrix $\boldsymbol{P}$ at row $i= xp+y+1$ and column $j =up+v+1$ for $(x, y, u, v) \in \llbracket 0, p-1\rrbracket^4$ is given by:
\begin{equation}
\boldsymbol{P}_{i, j} =  \frac{2}{p}  \alpha(u)\alpha(v) \cos\left[\frac{(2x + 1)u\pi}{2p}\right]   \cos\left[\frac{(2y + 1)v\pi}{2p}\right] 
\label{dctcoeff}
\end{equation}

\noindent where $\alpha(u)= \left\{
    \begin{array}{ll}
       \frac{1}{\sqrt{2}} & \mbox{if } u = 0 \\
        1 & \mbox{otherwise}
    \end{array}
\right.$  is a normalizing scale factor to make the transformation orthonormal. The columns of the matrix $\boldsymbol{P}$ are, in fact, the basis in which the signal is decomposed (see Fig. \ref{dctbasis}a). The matrix $\boldsymbol{P}$ is considered as a basis, in the sense that every signal (represented as a vector) can be decomposed in a unique way as a linear combination of the columns of $\boldsymbol{P} $. Alternatively, the term "dictionary" is also used in image processing. In the case of DCT, this basis has the particularity of being orthonormal, which implies that $\boldsymbol{P}^{-1} = \boldsymbol{P}^T$ where the superscript $T$ denotes the transpose operator. The elements of this basis are generally ordered, in the zig-zag pattern, from the smoothest vector to the one containing the highest frequencies (see Fig. \ref{dctbasis}a). This ordering is very useful for applications in compression as frequencies higher than a certain threshold are typically canceled.

A small improvement of \cite{DCT}, called adaptive aggregation and inspired from \cite{BM3D}, was proposed in \cite{multiscaleDCT}. The idea is to give higher weight to patches that have a sparser representation in the DCT domain, enabling to reduce the ringing effects near edges. The expression of the improved DCT denoiser then becomes: 
\begin{equation}
    F(\boldsymbol{y})_k = \frac{1}{W_k} \sum_{i = -q}^{q} \sum_{j = -q}^{q} w_{i, j, k} [\boldsymbol{P}  \varphi_{\lambda}  (\boldsymbol{P}^{-1} \boldsymbol{y}_{k, p}^{i,j})]_{s(i, j)}
\label{dctdenoiserv2}
\end{equation}

\noindent with $w_{i, j, k} =  (1 + \| \varphi_{\lambda}( \boldsymbol{P}^{-1} \boldsymbol{y}_{k, p}^{i,j}) \|_0)^{-1}$, where $\|.  \|_0$ denotes the $\ell_0$ pseudo-norm that counts the number of non-zero entries and $\displaystyle W_k = \sum_{i = -q}^{q}  \sum_{j = -q}^{q} w_{i, j, k}$. We used this latter expression to  derive  DCT2net.

\subsection{DCT2net: a CNN representation of a DCT denoiser}

\begin{figure*}[t]
\centerline{\includegraphics[scale=0.35]{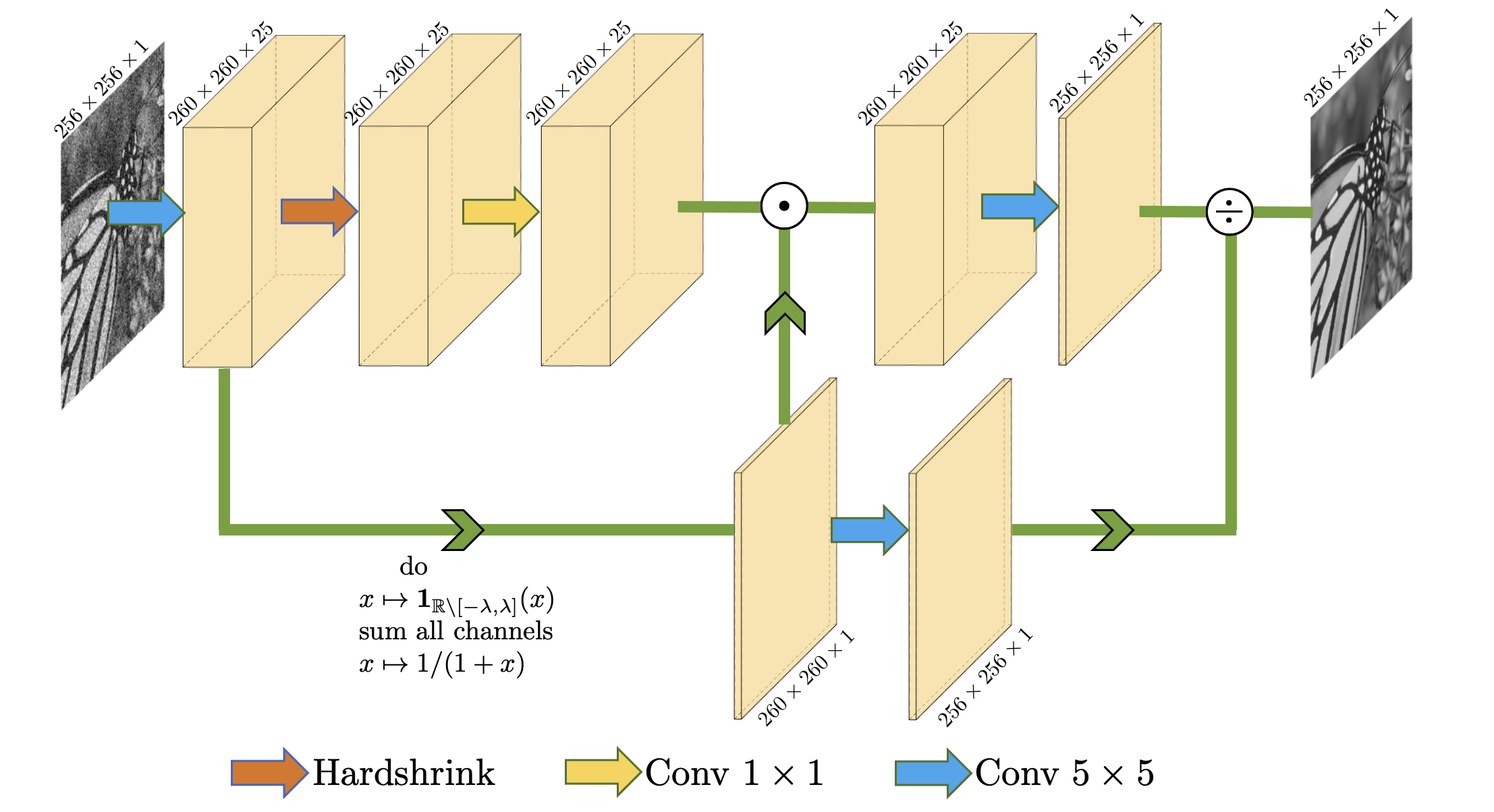}}
\caption{Architecture of DCT2net for a patch size $p=5$.}
 \label{DCT2net}
\end{figure*}

Interestingly, one of the easiest implementation of a DCT denoiser, as formulated in  (\ref{dctdenoiserv2}), can be done with a neural network. Indeed, all operations involved can be interpreted in terms of convolutions with a hard shrinkage function as activation function. 

Our shallow CNN is first composed of a convolutional layer with kernel size $p \times p$ leading to $p^2$ output channels. Figure \ref{DCT2net} shows such an architecture when $p = 5$. Note that the weights involved in kernels are to be found in the matrix $\boldsymbol{P}^{-1}$: each row is associated with one convolutional kernel. The action of this layer can be summed up as follows: each pixel is replaced by the patch formed by its neighborhood, after transformation by $\boldsymbol{P}^{-1}$. Then, the hard shrinkage function is applied element-wise. Afterwards, a $1 \times 1$ convolution layer operates on those patches where the weights correspond to the elements of the matrix $\boldsymbol{P}$. In order to compute the adaptive aggregation, we have to generate a weight map from the first layer computing the values $w_{i,j,k}$ in (\ref{dctdenoiserv2}). This latter is used to balance the features resulting from the second layer by channel-wise multiplication. 
The weighted pixels are then repositioned at their corresponding locations and then aggregated by summation. This can be implemented with the help of a last convolutional layer where the values of weights are either $0$ or $1$. Note that, for computational efficiency, a 2D transposed convolution is recommended. Finally, a normalization by $W_{k}$ (see (\ref{dctdenoiserv2})) is performed by dividing the last layer by the weight map, beforehand convolved by a kernel composed of ones.

We want to stress that our DCT2net is the strict implementation of the formulation in (\ref{dctdenoiserv2}). Equipped with the correct weights given by the definition of the DCT in (\ref{dctcoeff}), it exactly produces the same results as those obtained with the traditional implementation.

\subsection{Improvement of the transform}

As it is usually done with neural networks, we can train our DCT2net on an external dataset composed of $N$ pairs of noise-free and noisy images $(\boldsymbol{x}_i, \boldsymbol{y}_i)_{i \in \{ 1, \ldots, N\}}$ to improve the underlying transform. More precisely, our objective is to solve the following optimization problem:

\begin{equation}
    \boldsymbol{P}^* = \arg \min_{\boldsymbol{P}} \sum_{i=1}^{N} \|  F_{\boldsymbol{P}}(\boldsymbol{y}_i, \sigma_i) - \boldsymbol{x}_i \|^2_2
\label{equation4}
\end{equation}
\noindent where $F_{\boldsymbol{P}}$ denotes the network (Fig. \ref{DCT2net}) and $\boldsymbol{P}$ gathers the unknown parameters.

\begin{figure}[t]
\includegraphics[width=\columnwidth]{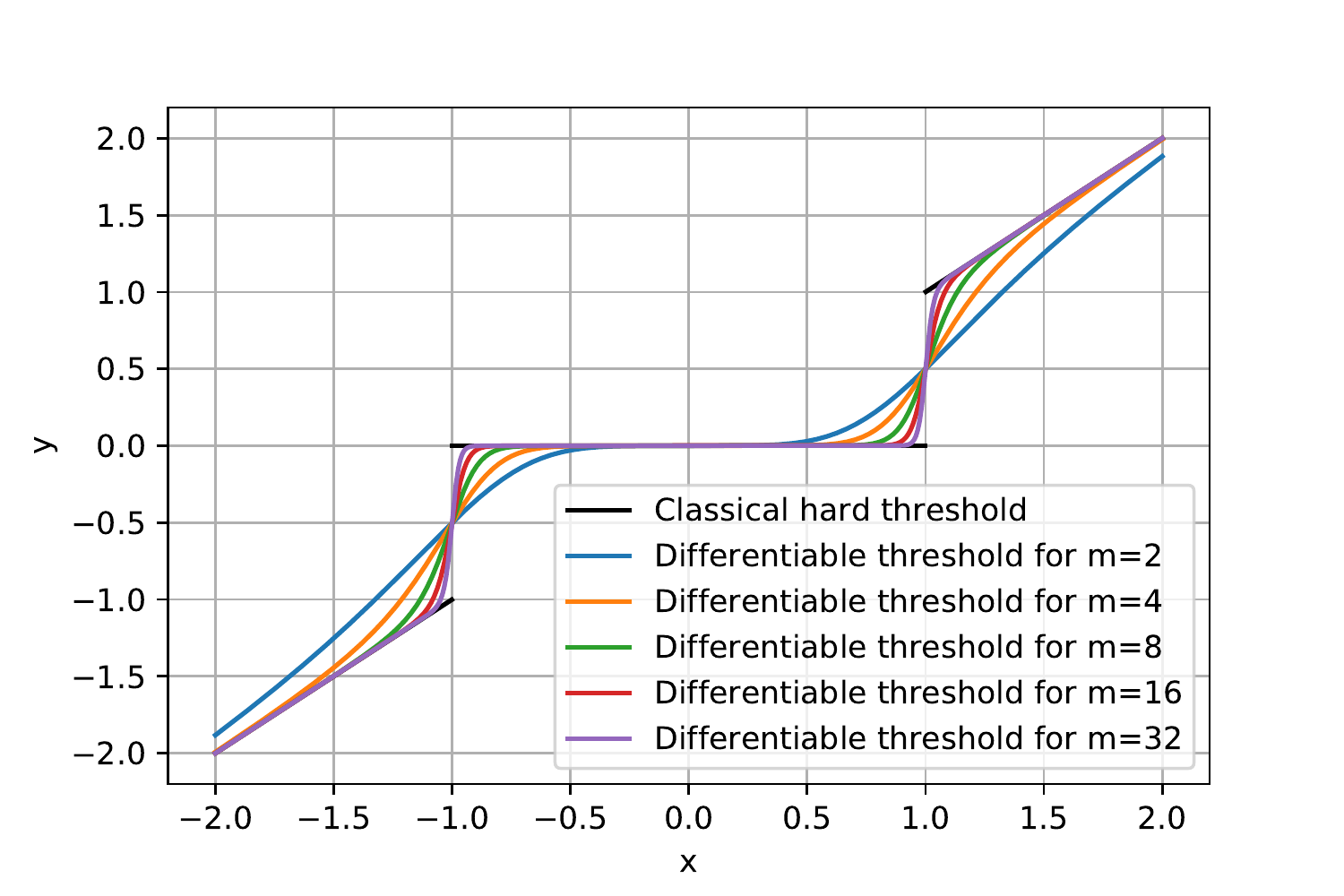} 
\caption{Approximation of the hard thresholing function $\varphi_{\lambda}$ by the sequence of differentiable functions $\varphi_{m, \lambda}$ for $\lambda=1$ and $m \in \{2, 4, 8, 16, 32\}$.}
\label{thresholdFig}
\end{figure}

To that extent, we need to restrict the model to the original transform where only one matrix is involved (the other one being its inverse) and where the other convolutions of the network composed of $0$ and $1$ are frozen. This can be achieved with the help of modern machine learning libraries such as Pytorch for which automatic differentiation can be kept on for complex operations such as matrix inversion but also deactivated for some layers. Nevertheless, the thresholding operation must be slightly adapted in a context of gradient descent where differentiation is needed. Thus, we replace the function $\boldsymbol{1}_{\mathbb{R}\backslash[-\lambda, \lambda]}$ by $\zeta_{m, \lambda}(x) = \frac{x^{2m}}{x^{2m} + \lambda^{2m}}$ with $m \in \mathbb{N}^*$. This choice is legitimate as the sequence of functions $(\zeta_{m, \lambda})_{m \in \mathbb{N}^*}$ converges pointwise to $\boldsymbol{1}_{\mathbb{R}\backslash[-\lambda, \lambda]}$. The hard shrinkage function $\varphi_{\lambda}$ then becomes $\varphi_{m, \lambda}(x) = \frac{x^{2m+1}}{x^{2m} + \lambda^{2m}}$. Figure \ref{thresholdFig} shows how close this approximation is from the original hard shrinkage function with ever growing values of $m$. By the way, this approximation is adopted only during the training phase for facilitating the optimization process. To stick with the original DCT denoiser, we use the original hard shrinkage function for the testing phase that gives the same results in terms of PSNR with no noticeable visual differences for the denoised images. It is worth noting that the use of the original hard shrinkage function instead of our differentiable approximation for the training phase does not work in our case, leading to a poor suboptimal local minimum, even though this activation function is available in most modern machine learning libraries. It is likely that this disappointing behavior is due to the discontinuity of the original function.

Finally, as recommended in \cite{DCT}, the threshold parameter $\lambda$ is set to $3 \sigma$ to significantly remove noise. But the choice of the multiplicative constant in front of $\sigma$ is actually of little importance and any other constant would produce same results as long as $\boldsymbol{P}$ can adapt itself.  Indeed, for two levels of threshold $\lambda$ and $\lambda'$, we have $\varphi_{\lambda}(x) = \frac{\lambda}{\lambda'} \varphi_{\lambda'}( \frac{\lambda'}{\lambda}x)$. In particular, for two choices of multiplicative constant $c$ and $c'$, $\varphi_{c \sigma}(x) = \frac{c}{c'} \varphi_{c' \sigma}( \frac{c'}{c}x)$, hence $\boldsymbol{P} \varphi_{c \sigma}(\boldsymbol{P}^{-1} \boldsymbol{y})  = \boldsymbol{Q} \varphi_{c' \sigma}(\boldsymbol{Q}^{-1} \boldsymbol{y})$ with $\boldsymbol{Q} = \frac{c}{c'} \boldsymbol{P}$ This means that, in theory, choosing a value other than $3$ would result in estimating the same transform, up to a multiplicative constant, and with exactly the same denoising performance. Appendix \ref{appendix2} gives more details about the choice of threshold, studying more particularly the case where multiple thresholds are used.

Note that, in practice, optimizing over the set of invertible matrices $\mathcal{GL}_{p^2}(\mathbb{R})$ in (\ref{equation4}) is not an issue and the problem can be treated through stochastic gradient descent without 
specific precaution. It is attributable to the fact that  $\mathcal{GL}_{p^2}(\mathbb{R})$ is dense in $\mathcal{M}_{p^2}(\mathbb{R})$ but $\mathcal{M}_{p^2}(\mathbb{R}) \backslash \mathcal{GL}_{p^2}(\mathbb{R})$ is not.

\section{A non-intuitive learned transform} \label{section3}

\begin{figure}[t]
\centering
\renewcommand{\arraystretch}{0.5}
\begin{tabular}{cc}
\includegraphics[scale=0.29]{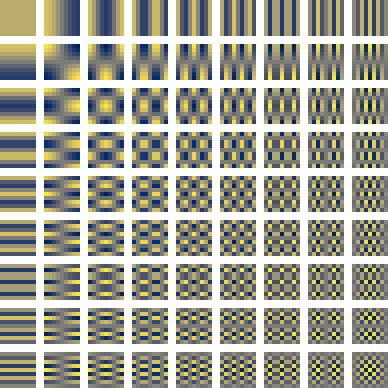} &
\includegraphics[scale=0.29]{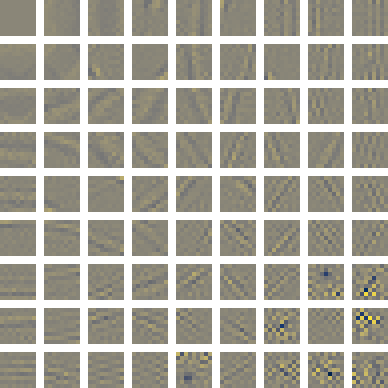} \\
\footnotesize (a) Original DCT basis & \footnotesize (b) DCT2net learned basis \\ 
\\

\includegraphics[scale=0.29]{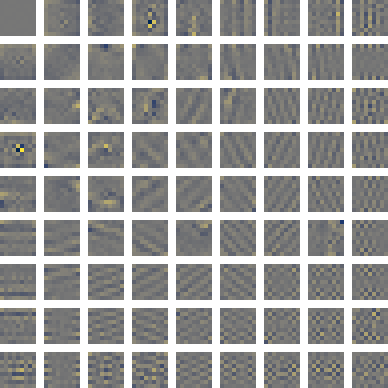} &
\includegraphics[scale=0.29]{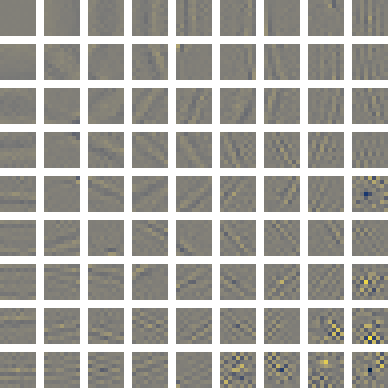} \\
\footnotesize (c) DCT2net trained on flat areas & \footnotesize (d) DCT2net trained on contours \\ 
\\

\includegraphics[scale=0.29]{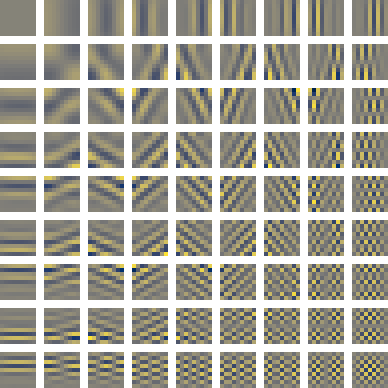} &
\includegraphics[scale=0.29]{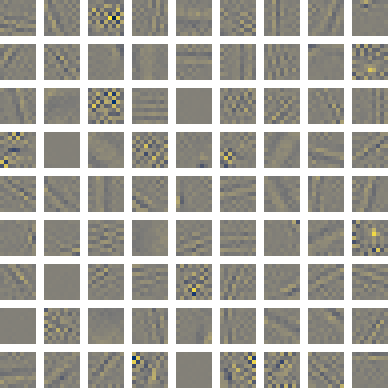} \\
\footnotesize   \begin{tabular}{c} (e) DCT2net trained to effectively \\  denoise patches before aggregation \end{tabular}  & \footnotesize   \begin{tabular}{c} (f) DCT2net with random \\   initialization\end{tabular} \\

\end{tabular}
\renewcommand{\arraystretch}{1}
\caption{Different bases in which patches are decomposed and thresholded for image denoising.}
\label{dctbasis}
\end{figure}

What is particularly attractive in our model is that we can easily display the learned transform and thus have a direct visual intuition of what the network has learned. Once DCT2net has been trained on an external dataset, a "pseudo" basis is derived, which is not orthonormal, but presumably more appropriate to encode non-stationary signals than the conventional DCT. Figure \ref{dctbasis} shows a visual comparison of the learned bases in different contexts for patches of size $9 \times 9$.

\subsection{On the orthonormality of the learned transform}

\addtolength{\tabcolsep}{-2pt} 
\addtolength{\tabcolsep}{2pt}

\addtolength{\tabcolsep}{-2pt} 
\begin{figure}[t]
\centering
\renewcommand{\arraystretch}{0.5}
\begin{tabular}{cc}
\includegraphics[scale=0.29]{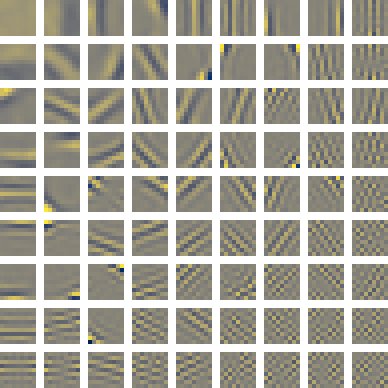} &
\includegraphics[scale=0.29]{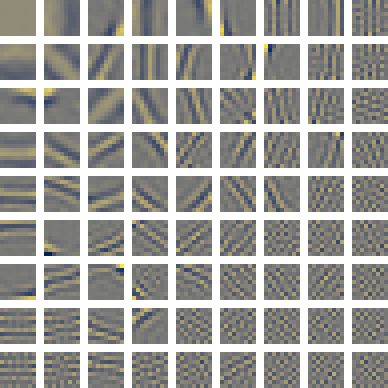}  \\
\footnotesize (a) $\lambda=3\sigma$ / 28.98 dB & \footnotesize (b) $\lambda=2.55\sigma$ / 29.28 dB    \\ 
\\
\includegraphics[scale=0.29]{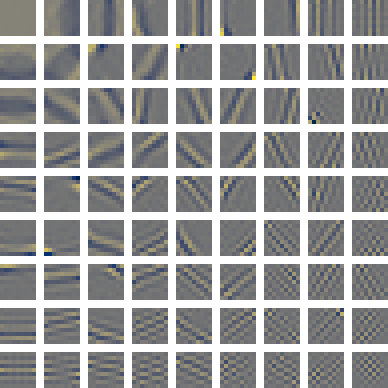} & \includegraphics[scale=0.29]{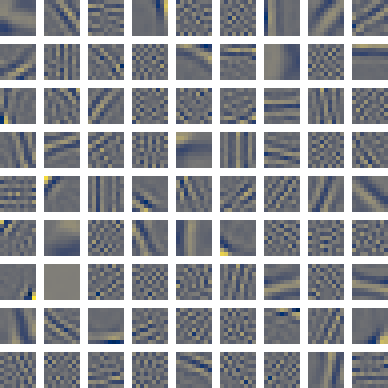} \\
\footnotesize \begin{tabular}{c}  (c)  Several thresholds / 29.44 dB \\ \\ \end{tabular} & \footnotesize    \begin{tabular}{c} (d) $\lambda=2.55\sigma$ / 29.20 dB  \\  (random initialization) \end{tabular}   \\
\end{tabular}
\renewcommand{\arraystretch}{1}

\caption{Orthonormal bases learned by DCT2net by addition of a regularization term. The threshold used is indicated (learned by optimization process when different from $3\sigma$) as well as the average PSNR on Set12 for $\sigma=25$ with each of these orthonormal bases. By way of comparison, the average PSNR on Set12 for the unconstrained DCT2net with patch size $9\times9$ is 29.57 dB. The basis exposed with several thresholds corresponds actually to an orthogonal basis with only one threshold as shown in appendix \ref{appendix3}.}
\label{dctbasis2}
\end{figure}
\addtolength{\tabcolsep}{2pt}

In the definition of DCT2net (\ref{dctdenoiserv2}), we impose no orthonormality constraint to learn the basis. Therefore, it is no wonder that, during the building of the matrix $\boldsymbol{P}$ through stochastic gradient descent, the property of orthonormality gets lost. One way \footnote{For a direct technique to derive an orthonormal matrix, with similar results compared to the regularization form, see appendix \ref{appendix1}.} to address this issue is to add a regularization term that encourages orthonormality  in the optimization process. The problem amounts to solving:

\begin{equation}
    \boldsymbol{P}^* = \arg \min_{\boldsymbol{P}} \sum_{i=1}^{N} \|  F_{\boldsymbol{P}}(\boldsymbol{y}_i, \sigma_i) - \boldsymbol{x}_i \|^2_2 + \beta \|I - \boldsymbol{P}^T\boldsymbol{P}\|_1 
\end{equation}
\noindent where $\beta \geq 0$ is  the regularization parameter. We consider here the $\ell_1$ norm, defined for a matrix $\boldsymbol{A}$ by $\|\boldsymbol{A}\|_1 = \sum |a_{i,j}|$, but the $\ell_2$ norm can be used instead. One can prove that this optimization problem with a penalty term corresponds to an underlying constraint problem of the form:

\begin{equation}
\begin{aligned}
\boldsymbol{P}^* =  \arg \min_{\boldsymbol{P}} & \sum_{i=1}^{N} \|  F_{\boldsymbol{P}}(\boldsymbol{y}_i, \sigma_i) - \boldsymbol{x}_i \|^2_2  \\
& \textrm{s.t.} \quad \|I - \boldsymbol{P}^T\boldsymbol{P}\|_1 \leq t\\
\end{aligned}
\end{equation}

\noindent which makes explicit the constraint of "close-orthonormality". Note that the parameter $t$ depends both on $\beta$ and on the data $(\boldsymbol{x}_i, \boldsymbol{y}_i)_{i \in \{1, \ldots, N\}}$.

It is worth noting that the resulting matrix $\boldsymbol{P}^{*}$ is not guaranteed to be orthonormal whatever the parameter $\beta$ is. To derive an orthonormal matrix from $\boldsymbol{P}^{*}$, we can select its nearest orthonormal matrix $\boldsymbol{P}_{\text{ortho}}$ in the Frobenius norm sense. The unique solution is given by $\boldsymbol{P}_{\text{ortho}} = \boldsymbol{U}\boldsymbol{V}^T$ where $\boldsymbol{U}$ and $\boldsymbol{V}$ are the matrices from the singular value decomposition of $\boldsymbol{P}^{*} = \boldsymbol{U} \boldsymbol{\Sigma} \boldsymbol{V}^T$. 

\begin{figure}[t]
\centering
\includegraphics[width=\columnwidth]{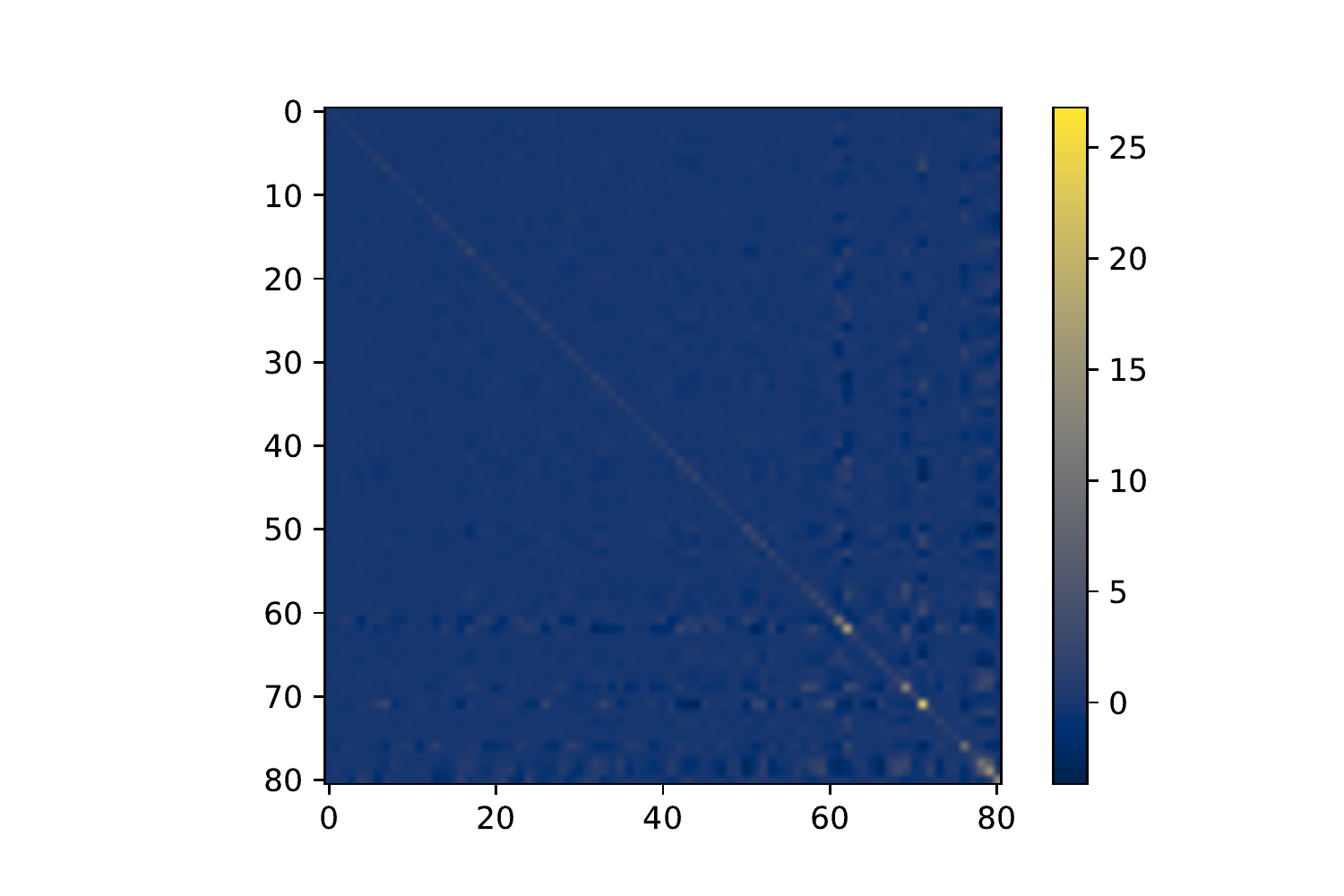} 
\caption{Representation of the matrix $\boldsymbol{P}^T \boldsymbol{P}$ where $\boldsymbol{P}$ denotes the transform learned by DCT2net for patches of size $9\times9$. If $\boldsymbol{P}$ were orthogonal, $\boldsymbol{P}^T \boldsymbol{P}$ would be equal to an invertible diagonal matrix. This is not strictly the case here, but we can notice that the elements outside the diagonal are very close to 0. Moreover, the diagonal represents an important weight of this matrix as $\sum_{i} q^2_{i,i} / \sum_{i, j} q^2_{i,j} \approx 60\%$ where the $q_{i,j}$ designate the coefficients of $\boldsymbol{P}^T \boldsymbol{P}$.}
\label{orthogonality}
\end{figure}

However, adding a constraint of orthonormality limits the expressivity of the network and we observed that the denoising performance was not as good as the regularization-free solution (see Fig. \ref{dctbasis2}). Moreover, the choice of the regularization parameter $\beta$ can be challenging as it needs to be adapted for each patch size. For all those reasons, we decided not to retain a solution with an orthonormal matrix. 
In spite of this, the matrix $\boldsymbol{P}$ learned by DCT2net is quite close to be orthogonal, even if no constraint has been imposed. This is illustrated in Fig. \ref{orthogonality} which displays the matrix $\boldsymbol{P}^T \boldsymbol{P}$ for  patches of size $9\times9$. We can notice that the non-diagonal elements are close to zero, which is expected for a matrix close to be orthogonal. In appendix \ref{appendix3}, we show how orthogonal and orthonormal matrices are linked, stating that if $\boldsymbol{P}$ is orthogonal, there exists an orthonormal matrix $\boldsymbol{Q}$ that would give exactly the same results in DCT2net as long as we set a different threshold by element of the basis.

Finally, as regards the initialization for the stochastic gradient descent, the original DCT basis (see (\ref{dctcoeff})) is considered by default. Considering random initializations such as Xavier initialization which is common in deep neural networks, lead to similar bases, even though the time for convergence is slightly more important. The convergence to the same solution whatever the initialization is a very good news, suggesting that optimizing the underlying non-convex problem is tractable. 

\subsection{DCT2net does not denoise patches}

\begin{figure}[t]

\begin{tikzpicture}
 
    \node (image) at (0,0) {
            \begin{tabular}{cc}
            
            \multirow{3}[3]{*}[1.00in]{\includegraphics[width=0.5\columnwidth]{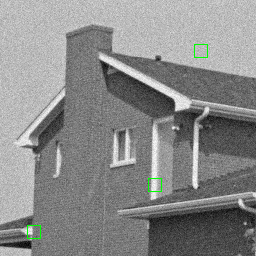}} & \includegraphics[width=0.4\columnwidth]{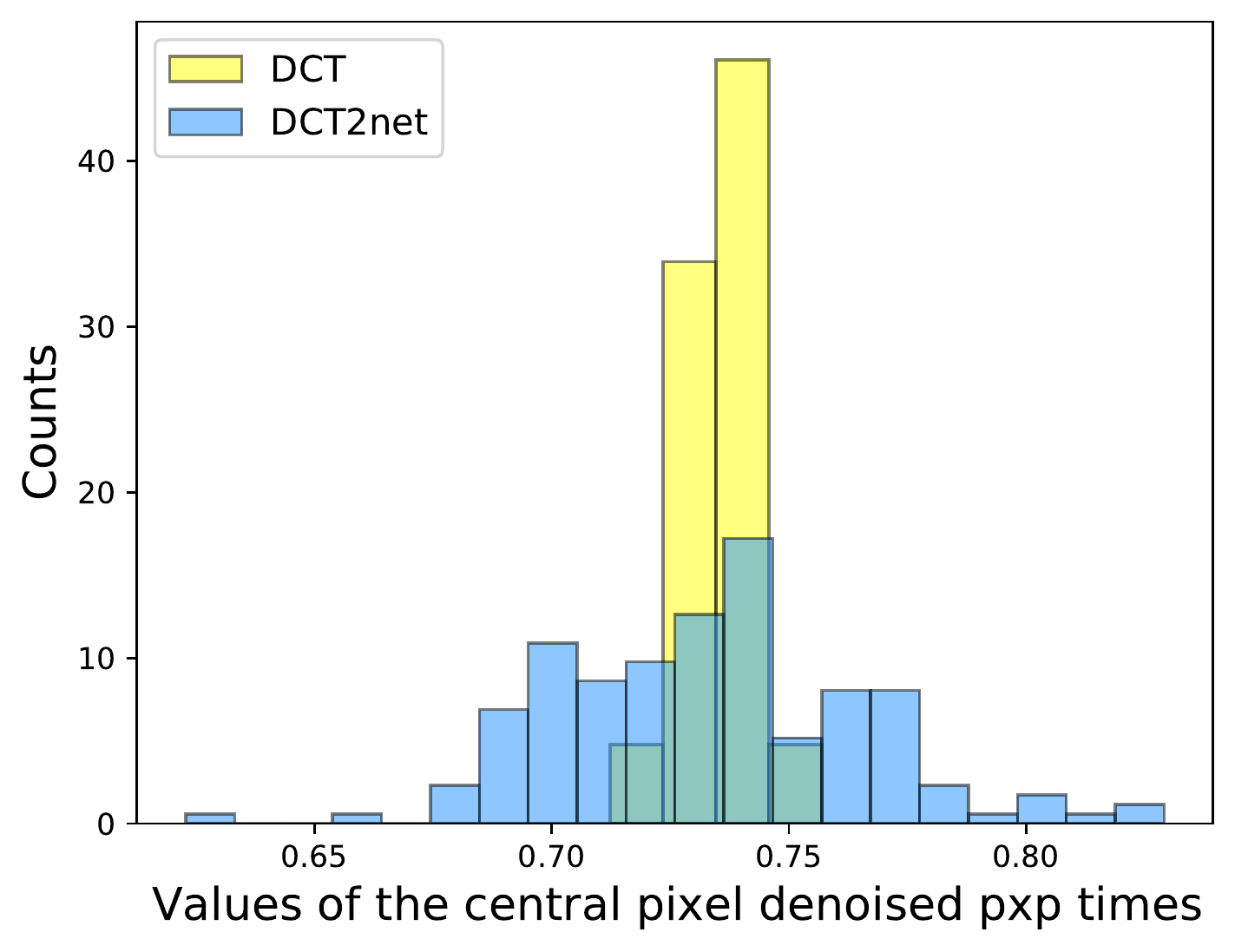} \\
            & \addtolength{\tabcolsep}{-5pt}\renewcommand{\arraystretch}{0.5}\begin{tabular}{ccc}
            \includegraphics[scale=0.3]{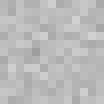} &
            \includegraphics[scale=0.3]{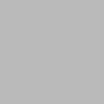} &
            \includegraphics[scale=0.3]{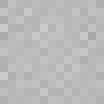} \\
            \scriptsize Noisy & \scriptsize DCT &  \scriptsize DCT2net
            \end{tabular}\addtolength{\tabcolsep}{5pt}\renewcommand{\arraystretch}{1} \\
            & \\
            
            \includegraphics[width=0.4\columnwidth]{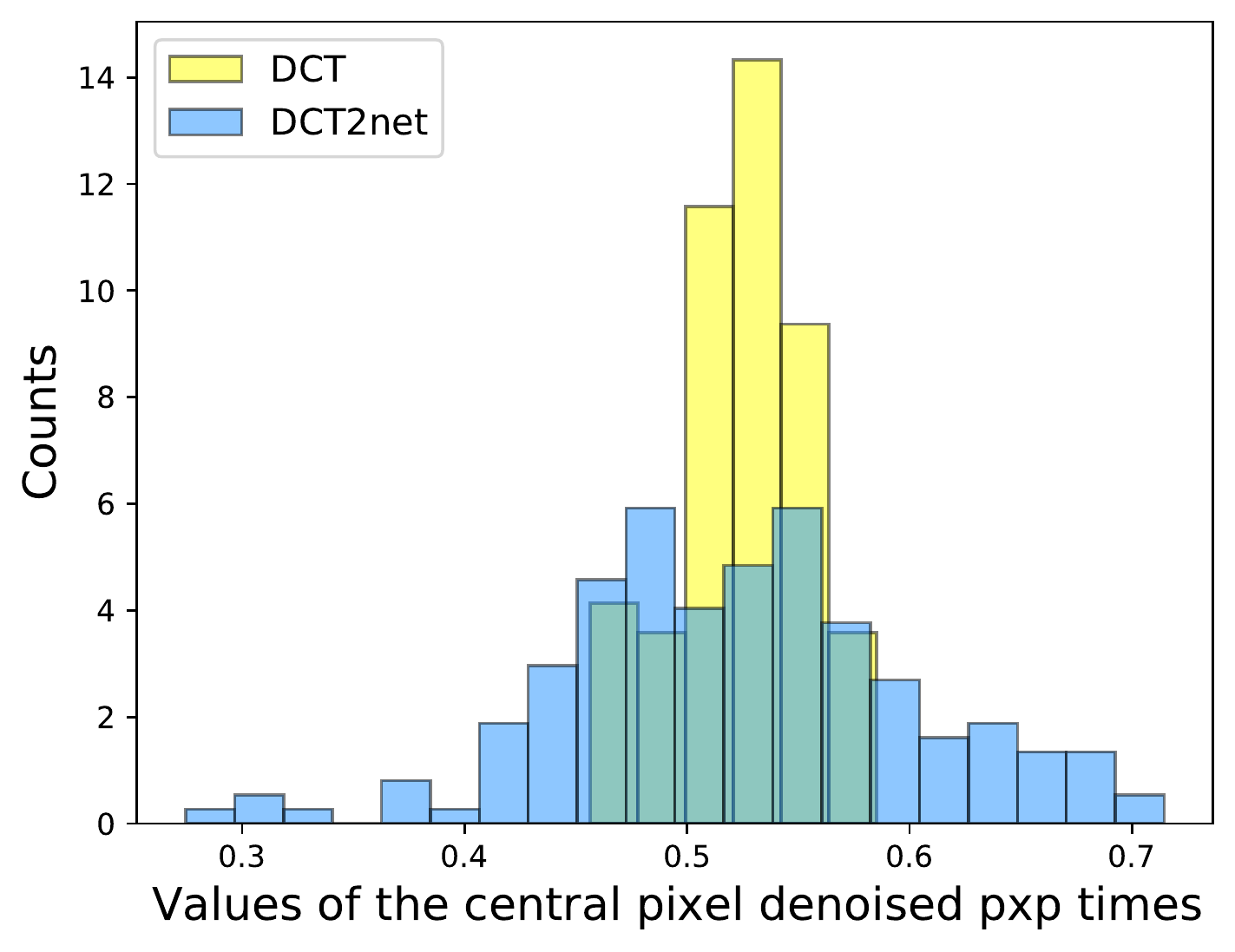} & \includegraphics[width=0.4\columnwidth]{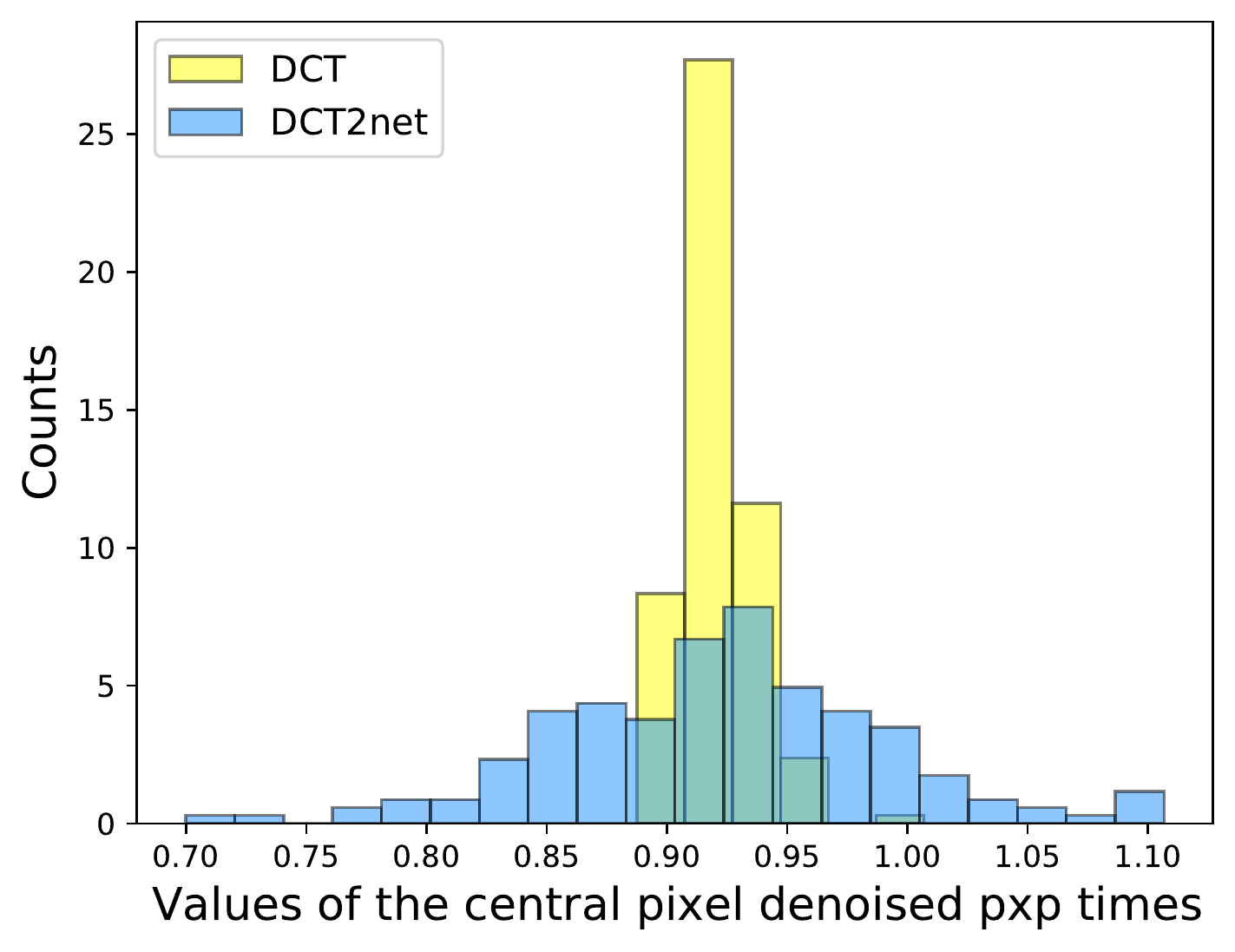} \\
            \addtolength{\tabcolsep}{-5pt} \renewcommand{\arraystretch}{0.5}
            \begin{tabular}{ccc}
            \includegraphics[scale=0.3]{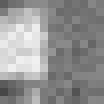} &
            \includegraphics[scale=0.3]{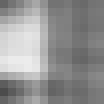} &
            \includegraphics[scale=0.3]{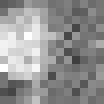} \\
            \scriptsize Noisy & \scriptsize DCT &  \scriptsize DCT2net
            \end{tabular}\addtolength{\tabcolsep}{5pt}\renewcommand{\arraystretch}{1}   & \addtolength{\tabcolsep}{-5pt}\renewcommand{\arraystretch}{0.5} \begin{tabular}{ccc}
            \includegraphics[scale=0.3]{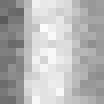} &
            \includegraphics[scale=0.3]{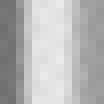} &
            \includegraphics[scale=0.3]{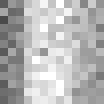} \\
            \scriptsize Noisy & \scriptsize DCT &  \scriptsize DCT2net
            \end{tabular}\addtolength{\tabcolsep}{5pt}\renewcommand{\arraystretch}{1}  \\
            
            \end{tabular}
            
    };
    
    \draw[ thick,green] (0.7,0.175) rectangle (4.25,4.55) node[below left,black,fill=green]{\small 1};
    
    \draw[ thick,green] (-3.75,-4.5) rectangle (-0.2,-0.1) node[below left,black,fill=green]{\small 2};
    
    \draw[ thick,green] (0.7,-4.5) rectangle (4.25,-0.1) node[below left,black,fill=green]{\small 3};

    \draw[ thick,green] (-0.8,3.58) -> (0.7,2.5);
    \draw[ thick,green] (-3.68,0.44) -> (-2,-0.1);
    \draw[ thick,green] (-1.45,1.38) -> (0.7,-0.1);
 
\end{tikzpicture}
\caption{For each noisy patch of size $13 \times 13$ extracted from \textit{House} image corrupted by AWGN with $\sigma=10$, its denoised version is displayed when processed by the original DCT and by the transform learned by DCT2net. The patches produced by DCT2net are very noisy compared to the original patches and patches denoised with DCT. Histograms show a comparison of the $p^2$ denoised values for the central pixel after transformation in each of the $p^2$ patches it belongs to ($p=13$). The variance of pixels intensities is higher with DCT2net.}
\label{patchDen2}

\end{figure}

When displaying the learned basis on a popular image dataset such as BSD400 \cite{berkeley}, one may be surprised. Interestingly, this basis, in which each patch is decomposed, is much more disorganized than the original DCT basis. One may doubt that the DCT2net basis denoises better patches as it contains no clear pattern. As a matter of fact, applying this basis does not denoise patches but rather degrades them even more as shown in Fig. \ref{patchDen2}. The main reason is that the network actually denoise image patches and performs aggregation at once, making it difficult to understand why such a basis improves the PSNR value of the restored image.

We observed that, for a given noisy pixel $k$ belonging to $p^2$ patches, the $p^2$ "denoised" versions of pixel $k$ with DCT2net  have a very high variance when compared to the $p^2$ denoised values obtained with the  DCT denoiser, for which all $p^2$ denoised values are generally almost all the same, as illustrated in  Fig. \ref{patchDen2}. Nevertheless, after the adaptive aggregation step, the pixels denoised by applying the learned transform are closer, in average, to the ground truth ones. This is a counter-intuitive result that questions our preconceptions on denoising. This suggests that the final aggregation step is not a basic post-processing step but plays an important role in denoising, as confirmed below.

\begin{figure*}[t]
\centering
\begin{tabular}{ccccc}
\includegraphics[scale=0.35]{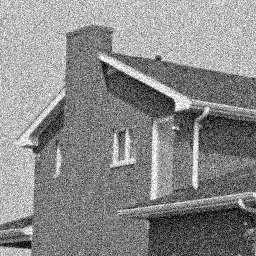} &
\includegraphics[scale=0.35]{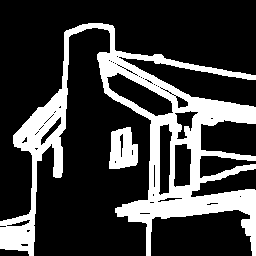} &
\includegraphics[scale=0.35]{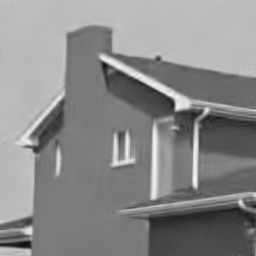} &
\includegraphics[scale=0.35]{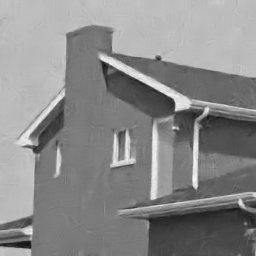} &
\includegraphics[scale=0.35]{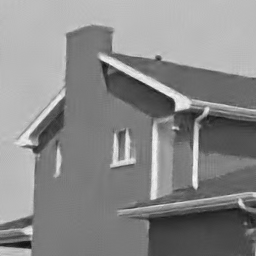}\\
\footnotesize Noisy / 20.17 dB & \footnotesize Binary classification map &  \footnotesize  DCT / 31.18 dB & \footnotesize DCT2net / 32.20 dB & \footnotesize DCT/DCT2net  / 32.26 dB \\

\end{tabular}

\caption{Application of the proposed procedure to get a partition of the noisy image \textit{House}. Pixels belonging to white areas after classification are denoised with DCT2net and the others with a traditional DCT denoiser. Results (in PSNR) are given for each method for a noise level $\sigma = 25$.}
\label{segmentation}
\end{figure*}

\subsection{Constraining DCT2net to effectively denoise patches is an unsuccessful strategy}

\begin{table}[t]
\centering
  \caption{The average PSNR (dB) results of two different transforms on patches of size $15\times 15$ of Set12 corrupted with white Gaussian noise and $\sigma=15$, $25$ and $50$.} \label{tablePatch}

  \begin{tabular}{*{4}{|c}|}
  \hline
  
  \rowcolor{lightgray}  \textbf{Methods} & $\sigma = 15$ & $\sigma = 25$ & $\sigma = 50$  \\\hline\hline
  \multicolumn{4}{|c|}{\textbf{Before aggregation}} \\ \hline
  DCT & 27.74 & 25.19 & 21.92  \\\hline
  DCT2net trained on patches& 28.18 & 25.98 & 22.97  \\\hline
  \multicolumn{4}{|c|}{\textbf{After adaptive aggregation}} \\ \hline
  DCT & 31.08 & 28.53 & 25.37  \\\hline
  DCT2net trained on patches& 30.90 & 28.66 & 25.65  \\\hline
\end{tabular}
\end{table}

As the strategy followed by DCT2net is counter-intuitive and hardly comprehensible for a human brain, we tried to constrain the learned transform to effectively denoise patches. In this study, aggregation is performed in a second step with conventional weighted averages. This can be done in practice by cutting our neural network represented in Fig. \ref{DCT2net} after the two first convolutional layers, so that the output $F_{\boldsymbol{P}}(\boldsymbol{y}_i)$ is of size $(H-p+1) \times (W-p+1) \times p^2$. For a clean image $\boldsymbol{x}_i$, we can compute its patch representation $\Pi(\boldsymbol{x}_i)$ of size  $(H-p+1) \times (W-p+1) \times p^2$ as well and solve the following optimization problem: 

\begin{equation}
    \boldsymbol{P}^* = \arg \min_{\boldsymbol{P}} \sum_{i=1}^{N} \|  F_{\boldsymbol{P}}(\boldsymbol{y}_i, \sigma_i) - \Pi(\boldsymbol{x}_i) \|^2_2.
\end{equation}

Figure \ref{dctbasis}e shows the learned transform $\boldsymbol{P}^*$ that effectively denoise patches. The resulting matrix $\boldsymbol{P}^*$ is much more natural and interestingly very close the original DCT basis. The gain in PSNR on patches of this new transform was evaluated on the Set 12 dataset. The results reported in Table \ref{tablePatch} show that the learned transform produces systematically a higher PSNR in average for the patches than the traditional DCT. One could expect that this transform would outperform DCT after the aggregation step. Unfortunately, this is not the case. 
Once the transform has been re-used in our DCT2net shown in Fig. \ref{DCT2net} with the adaptive aggregation integrated, the expected boost compared to traditional DCT in terms of PSNR significantly decreases. The difference of PSNR is insignificant, with a visually less attractive result. We can notice an interesting phenomenon for $\sigma=15$: from better denoised patches, our learned transform fails to outperform the traditional DCT after any classical aggregation technique. This counter-intuitive result confirms, once again, that the aggregation step is equally important as  denoising  patches.

 To go beyond  DCT, the key issue is to follow a non-intuitive path that consists in degrading the patches and performing  aggregation step to rearrange everything and produce a high-quality denoised image.

\section{DCT2net mixed with DCT to reduce unpleasant visual artifacts} \label{section4}

\begin{figure}[t]
\centering
\begin{tikzpicture}
\node (node1) {\includegraphics[scale=0.25]{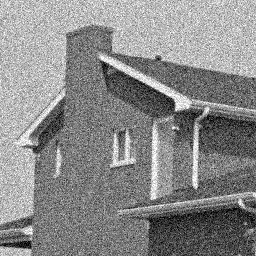}};
\node[below=0.3cm of node1] (node2) {\includegraphics[scale=0.25]{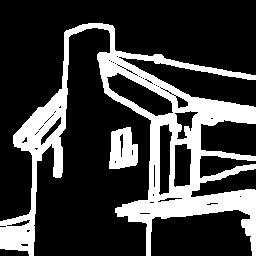}};
\node[above=-0.2cm of node1] (text1) {\small \textbf{Input noisy image}};

\node[left=0.69cm of node2] (node3) {\includegraphics[scale=0.25]{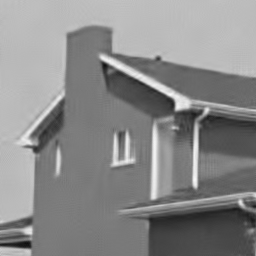}};
\node[right=0.69cm of node2] (node4) {\includegraphics[scale=0.25]{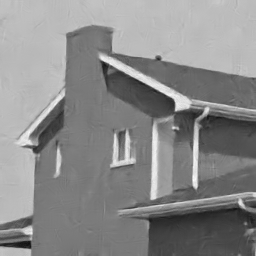}};

\node[above=-0.2cm of node3] (text2) {\small \textbf{DCT denoiser}};
\node[above=-0.2cm of node4] (text3) {\small \textbf{DCT2net denoiser}};

\node[above right=-1.1cm and -0.08cm of node3] (text4) {\tiny Binary};
\node[above right=-1.2cm and -0.24cm of node3] (text4) {\tiny classification};

\node[below left=0.1cm and -0.7cm of node2] (nodeNot) {\Large $\lnot$};
\node[below left=0.5cm and -0cm of node2] (node5) {\Large $\odot$};
\node[below right=0.5cm and -0cm of node2] (node6) {\Large $\odot$};

\node[below=0.5cm of node5] (node7) {\includegraphics[scale=0.25]{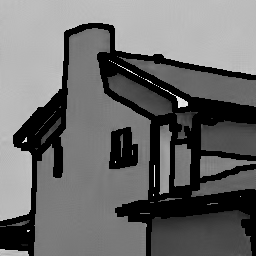}};
\node[below=0.5cm of node6] (node8) {\includegraphics[scale=0.25]{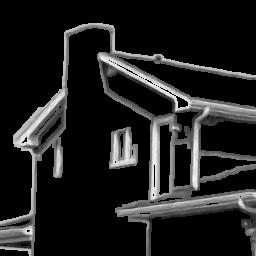}};

\node[below=7.4cm of node1] (node9) {\Large $\oplus$};

\node[below=0.5cm of node9] (node10) {\includegraphics[scale=0.25]{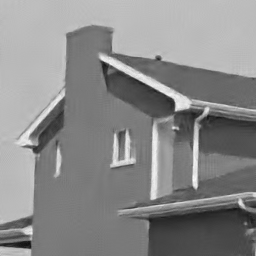}};
\node[below=-0.2cm of node10] (text10) {\small \textbf{Reconstructed image}};

\draw[->, ultra thick, mycolor, to path={-| (\tikztotarget)}] (node1) edge (text2) (node1) edge (text3);

\draw (node4) edge[->, ultra thick, mycolor] (node6);
\draw (node2) edge[->, ultra thick, mycolor] (node6);

\draw (node3) edge[->, ultra thick, mycolor] (node5);
\draw (node2) edge[->, ultra thick, mycolor] (node5);

\draw (node5) edge[->, ultra thick, mycolor] (node7);
\draw (node6) edge[->, ultra thick, mycolor] (node8);

\draw (node7) edge[->, ultra thick, mycolor] (node9);
\draw (node8) edge[->, ultra thick, mycolor] (node9);

\draw (node9) edge[->, ultra thick, mycolor]  (node10) ;

\draw (node3) edge[->, ultra thick, mycolor] (node2);
\end{tikzpicture}
\caption{Hybrid denoising scheme. A first rough denoising is performed by DCT to improve the classification procedure. Then the denoised image is recycled on the flat areas while DCT2net denoises the contours.}
\label{schema}
\end{figure}
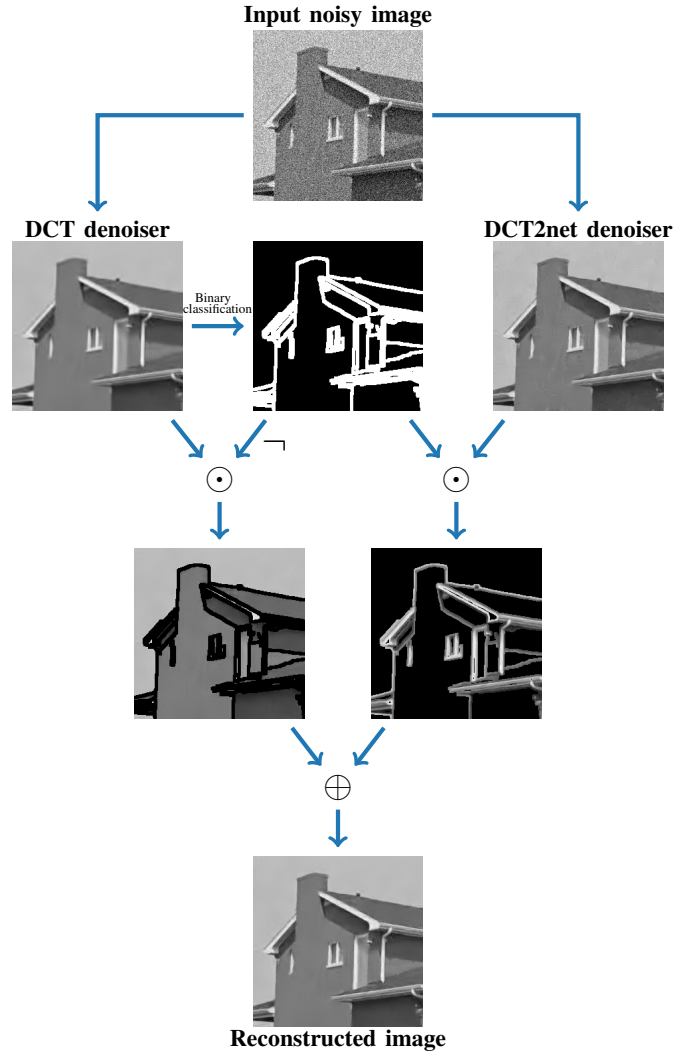

\begin{figure}[t]
\centering
\stackinset{r}{-0.0001\textwidth}{t}{-0.0001\textwidth}
  {\includegraphics[scale=0.3]{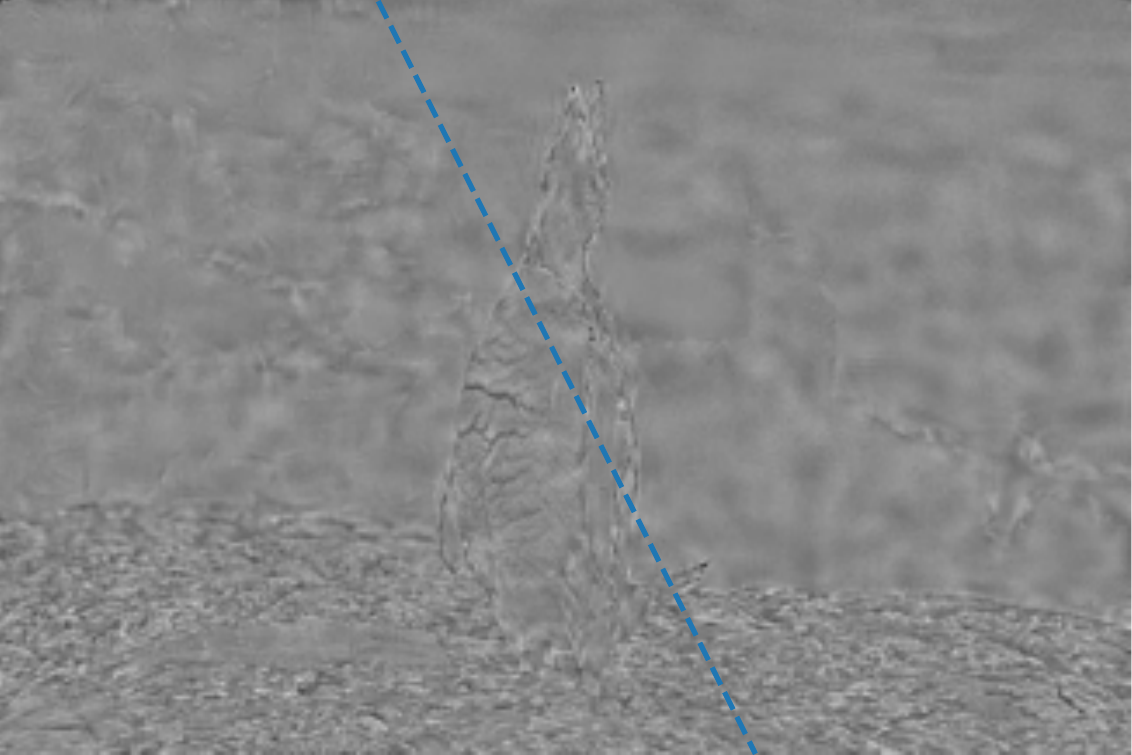}}
  {\includegraphics[width=\columnwidth]{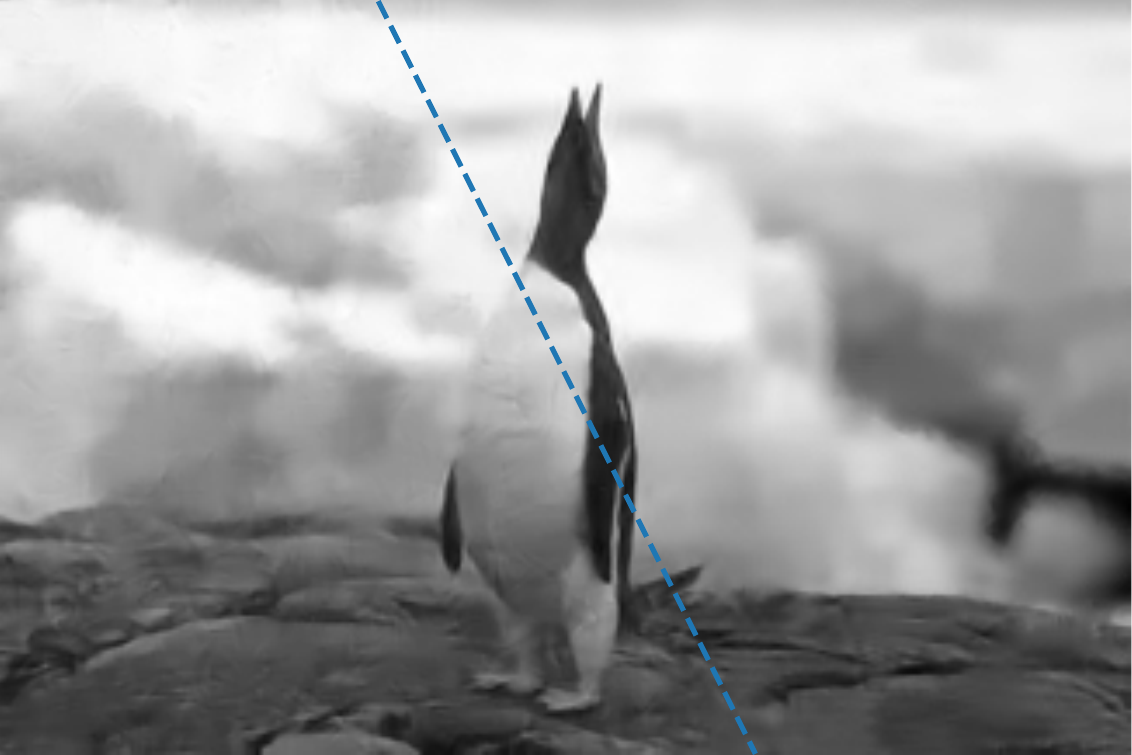}}
\caption{Image from BSD68 denoised with DCT2net producing unpleasant visual artifacts (left) and denoised using both DCT and DCT2net in collaboration (right) for $\sigma = 25$. The difference between the noise-free image and the denoised images is also provided, highlighting these artifacts.}
\label{showArtifacts}
\end{figure}

Even though  DCT2net produces high PSNR values as BM3D \cite{BM3D}, the visual results can be surprisingly not as good as expected, especially in flat regions in images. This is due to the emergence of structured unpleasant artifacts in those regions that are extremely eye-catcher. Figure \ref{showArtifacts} shows an example of those artifacts that are inherent to our method. They are difficult to characterize and very different from what we would get with a non-adaptive DCT denoiser. The visual impression is as if the recovered image had been scratched in several locations. These undesirable artifacts were probably promoted by blind stochastic gradient descent, to produce the best PSNR value in average. It is likely that this  blind choice originates from a trade-off between denoising flat regions and textures. The problem was not solved by considering an adequate loss function  \cite{loss} or by adopting a multi-scale scheme  \cite{multiscaleDCT}.

\addtolength{\tabcolsep}{-5pt}  
\begin{figure*}[t]
\centering
\begin{tabular}{cccccc}
\includegraphics[width=0.33\columnwidth]{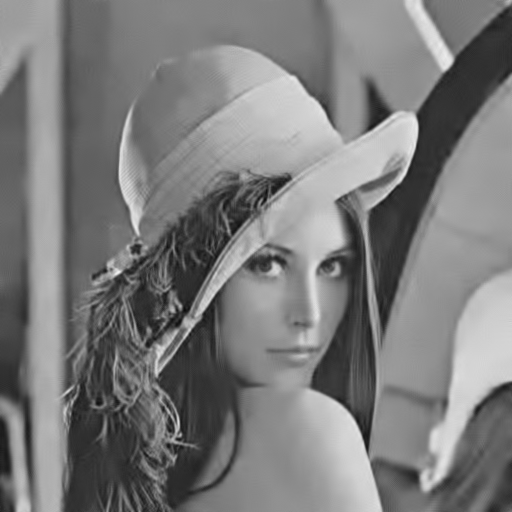} & \includegraphics[width=0.33\columnwidth]{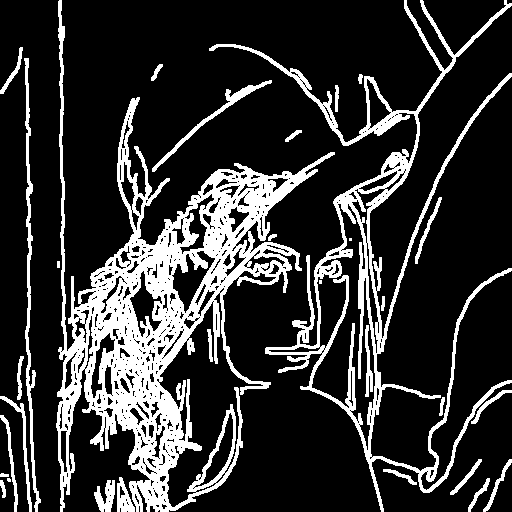} & \includegraphics[width=0.33\columnwidth]{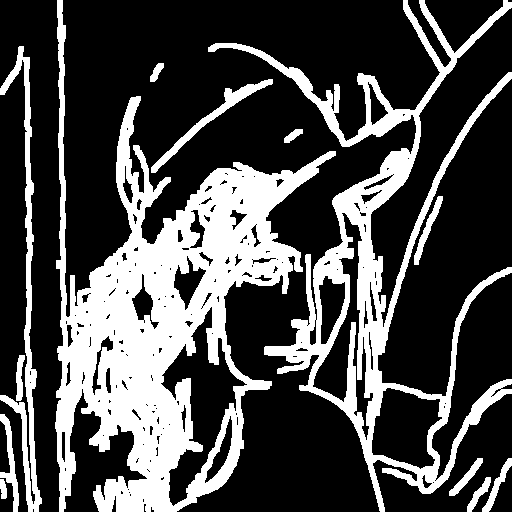} & \includegraphics[width=0.33\columnwidth]{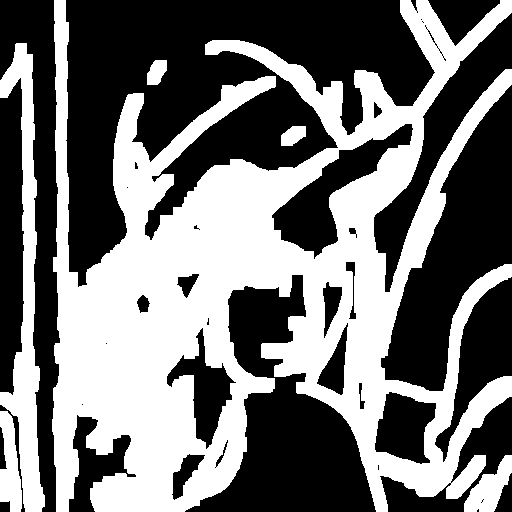} & \includegraphics[width=0.33\columnwidth]{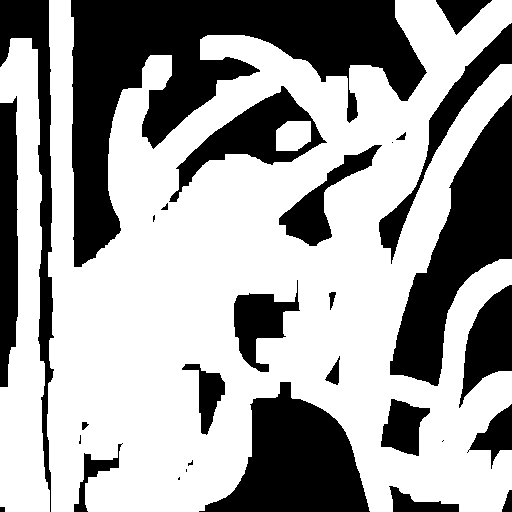}  & \includegraphics[width=0.33\columnwidth]{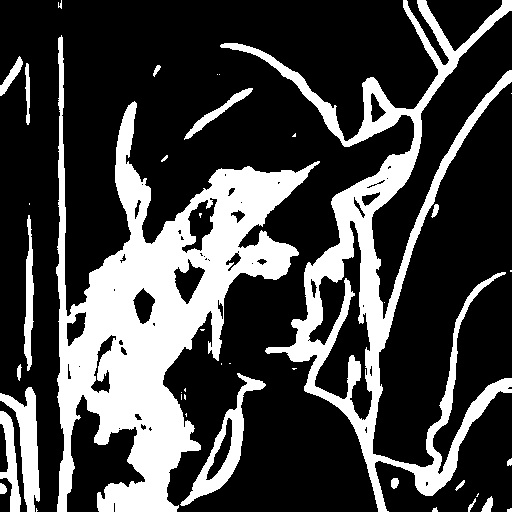} \\

\scriptsize DCT / 31.91 dB & 
\scriptsize \begin{tabular}{c} Canny mask\\ with dilation $3\times3$  \end{tabular}&
\scriptsize \begin{tabular}{c} Canny mask\\ with dilation $5\times5$  \end{tabular}&
\scriptsize \begin{tabular}{c} Canny mask\\ with dilation $11\times11$  \end{tabular}&
\scriptsize \begin{tabular}{c} Canny mask\\ with dilation $21\times21$  \end{tabular}&
\scriptsize \begin{tabular}{c} TV mask\\ with threshold of 75\%  \end{tabular}   \\

\includegraphics[width=0.33\columnwidth]{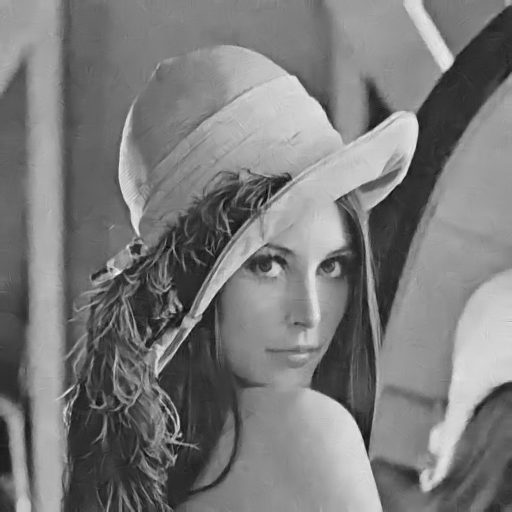} & \includegraphics[width=0.33\columnwidth]{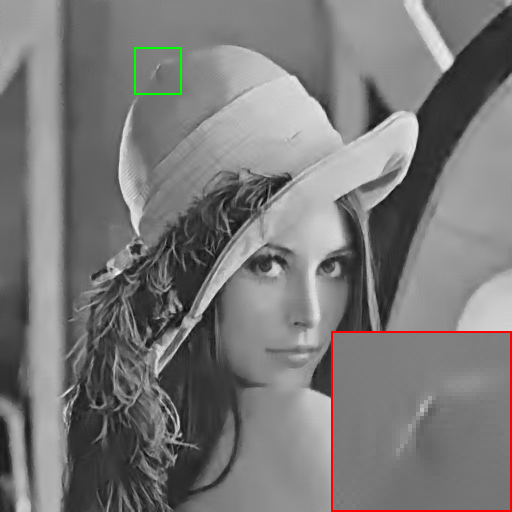} & \includegraphics[width=0.33\columnwidth]{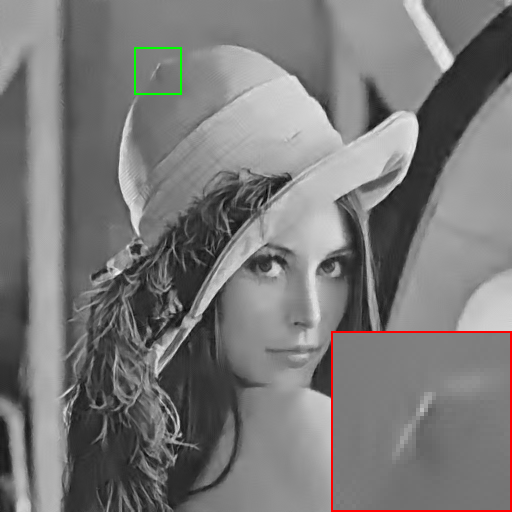} & \includegraphics[width=0.33\columnwidth]{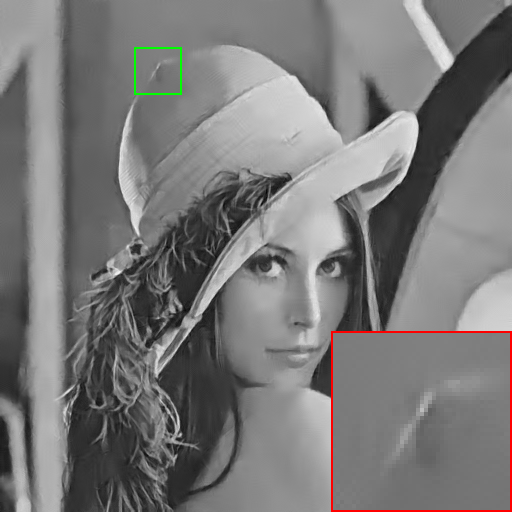} & \includegraphics[width=0.33\columnwidth]{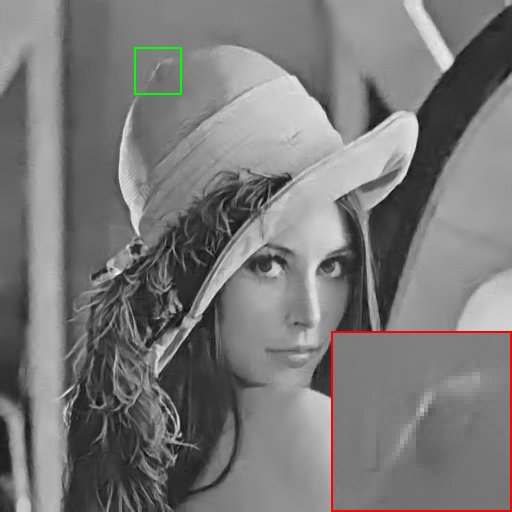}  & \includegraphics[width=0.33\columnwidth]{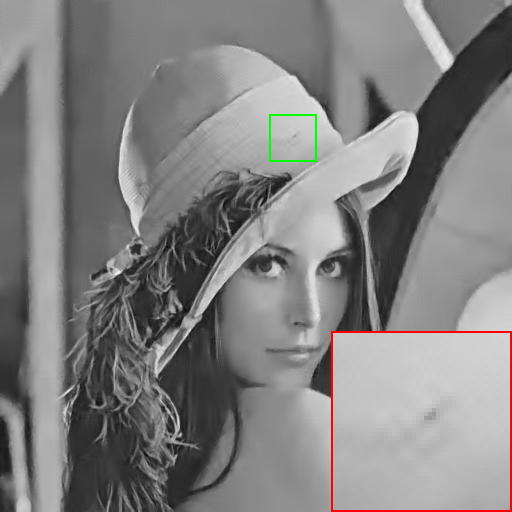} \\

\scriptsize DCT2net / 32.70 dB & 
\scriptsize DCT/DCT2net / 32.59 dB & 
\scriptsize DCT/DCT2net / 32.67 dB & 
\scriptsize DCT/DCT2net / 32.67 dB & 
\scriptsize DCT/DCT2net / 32.69 dB & 
\scriptsize DCT/DCT2net / 32.62 dB \\

\end{tabular}
\caption{Some examples of the classifications based on Canny edge detector and Total Variation on noisy \textit{Lena} for $\sigma = 20$.}
\label{cannyseg}
\end{figure*}
\addtolength{\tabcolsep}{5pt}

To tackle this problem, we  combine below the performance of both transforms, that is the original DCT and the transform learned by DCT2net. While the original DCT performs very well in flat regions, DCT2net recovers details more efficiently in the vicinity of contours and in fine textured regions. Our idea is then to classify the pixels into two classes and to apply the most appropriate denoiser at each pixel. In what follows, we show how a  binary map can be obtained, telling us which pixels are to be denoised with DCT2net or DCT. Figure \ref{segmentation} illustrates an example of such a procedure applied to the \textit{House} image.

The noisy image must be roughly denoised beforehand in order to robustly detect the flat regions, textured regions and contours. The DCT denoiser (\ref{dctdenoiserv2}) is appropriate in our case, as illustrated in Figure \ref{schema}. It has the advantage of being particularly cost-efficient and nearly parameter-free. The resulting denoised image is also re-used to produce the final image, as illustrated in Fig. \ref{schema}, saving time and resources. Interestingly, letting the traditional DCT denoiser operate on a large majority of pixels of the noisy image does not alter the PSNR values in our experiments. As DCT produces smooth images in homogeneous regions, the artifacts are removed and the visual result is enhanced considerably.

\bigskip

 \textit{a) Classification based on Canny edge detector:} Multiple choices of classification techniques are possible to separate the flat regions from the contours and textured areas. A high precision in the classification is not required as our goal is  to isolate parts of the image that are susceptible to contain artifacts after denoising with DCT2net. 
 
 The  classification problem into two classes can be achieved with an efficient traditional technique:  the Canny edge detector \cite{canny}. This method uses Sobel filters in both horizontal and vertical direction at its core to compute the gradient for each pixel.  The direction of edges is then analyzed to remove any unwanted pixels which may not constitute contours. Finally, an hysterisis thresholding is used to decide which pixels, detected positively in the first instance, are actually edges. This last step is based on the spatial analysis of connectivity, ensuring some coherence in the final classification map. To enlarge the support of edges found by this Canny edge detector, we apply a simple dilation operation in the end.

In practice, the image is preliminary slightly smoothed with a unit standard deviation Gaussian filter. The lower and upper thresholds involved in the Canny edge detector are set respectively to $0.1$ and $0.2$ (values of pixels being in the interval $[0, 1]$). These thresholds are set once and for all and are not changed in all experiments. Finally, the dilation operation is performed using a kernel of size $5 \times 5$. This choice of size of dilation kernel was motivated by its good performance in terms of PSNR, without scarifying the visual quality depending on the amount of artifacts (which is the case for larger sizes). Table \ref{dilation} reports the influence of the size of the dilation operation on the PSNR values for $\sigma=20$ on the Set 12 dataset composed of $12$ widely used images for denoising. Unsurprisingly, the larger the dilation filter  (that is, the more pixels are processed with DCT2net), the higher the PSNR value is. It can be noticed that considering small or large kernels does not affect the PSNR values significantly. If the dilation is very large, it amounts to applying DCT2net to all pixels in the image.

\begin{table}[t]
\centering
  \caption{The average PSNR (dB) results of our DCT/DCT2net method on Set12 corrupted with white Gaussian noise and $\sigma=20$ for different sizes of dilation kernel.} \label{dilation}

  \begin{tabular}{*{7}{|c}|}
  \hline
  
  \rowcolor{lightgray}  Size of dilation & 3 & 5 & 7 & 9 & 11 & $\infty$  \\\hline\hline
  DCT/DCT2net & 30.58 & 30.67 & 30.69 & 30.70 & 30.71 & 30.75  \\\hline

\end{tabular}
\end{table}

\bigskip

\textit{b) Classification based on Total Variation:} Alternatively, classification can be performed by applying the local Total Variation (TV). This amounts to computing the sum of gradients on small windows, which is expected to be low in flat regions and high on edges. After computing the value of the TV for every pixel, a local-TV map is derived where values are all the more important as the original noisy pixel belongs to a complex geometry, that is edges or texture. By defining an arbitrary threshold, it is possible to partition the image into two distinct components: \textit{high gradient} and \textit{low gradient} pixels where DCT2net and DCT are applied, respectively.

\bigskip

\textit{c) Comparisons of classification methods:} Figure \ref{cannyseg} shows the computed binary masks on the image \textit{Lena} for the two aforementioned classification techniques as well as the recomposed denoised image using both DCT2net/DCT denoiser.

We can notice that the technique based on the Canny edge detector gives a coherent classification where almost all contours are detected. When dilating more and more those contours with larger and larger dilation kernels, the PSNR improves as more pixels are processed with DCT2net. However, there is a risk of generating unpleasant artifacts near contours as it is the case in our example with a dilation kernel of size $21\times21$ (see Fig. \ref{cannyseg}). That is why, we set the size of this kernel to $5\times5$ once and for all which is a good balance between performance based on PSNR value and subjective visual perception.

Compared to the classification based on the Canny edge detector, the TV-based one is more limited. Indeed, some edges are missing. Worse still, some isolated white blocks appear in the background. It is very troublesome as those isolated zones will create further unpleasant artifacts at the end, as shown on Figure \ref{cannyseg}. It appears that the \textit{denoising styles} of those two denoisers are not compatible on similar zones. The human eye quickly notices a lack of coherence in the \textit{denoising tone} which is prejudicial to the visual quality. We observed similar issues for all size of windows for the TV computation and for all thresholds. Those critical drawbacks are prohibitive in the application of a such a method and we decided to focus on the Canny edge detector  in our experiments.

\addtolength{\tabcolsep}{-5pt}    
\begin{figure*}[ht]
\centering
\begin{tabular}{cccccc}
\includegraphics[scale=0.26]{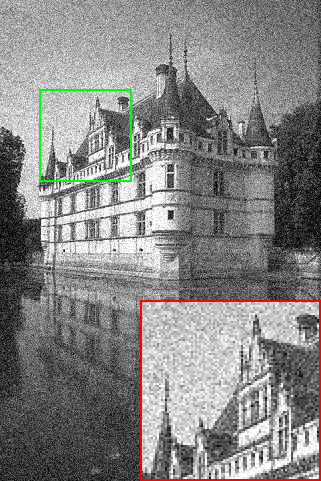} &
\includegraphics[scale=0.26]{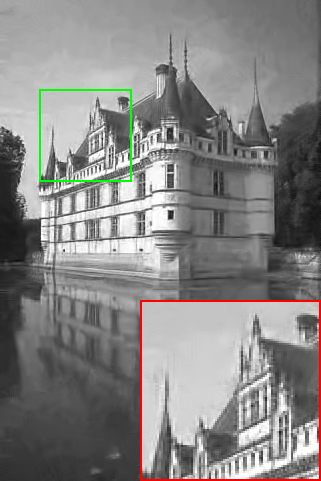} &
\includegraphics[scale=0.26]{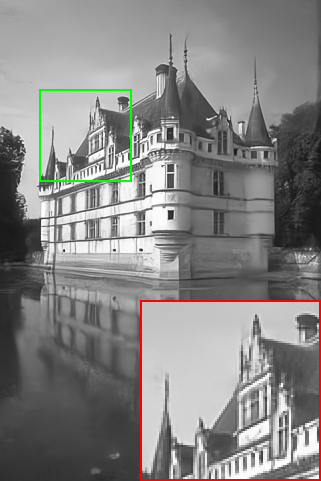} &
\includegraphics[scale=0.26]{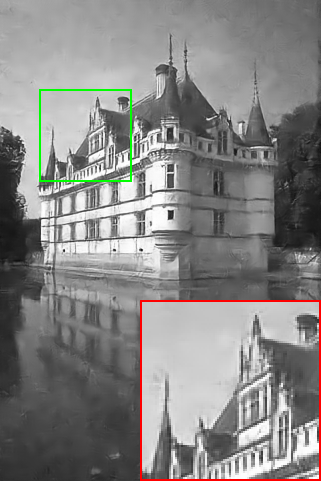} &
\includegraphics[scale=0.26]{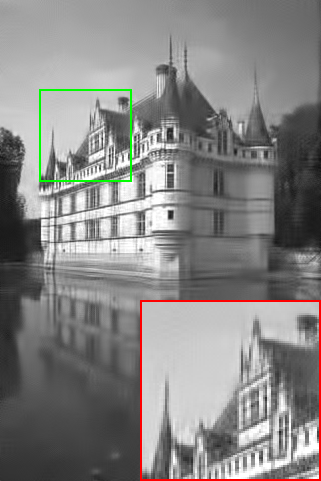} &
\includegraphics[scale=0.26]{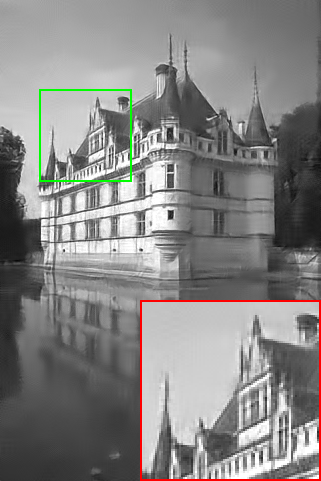}  \\
 \footnotesize Noisy / 20.17 dB  &   \footnotesize BM3D / 29.49 dB  &    \footnotesize DnCNN / 30.15 dB  & \footnotesize LKSVD$_{1,8,256}$ / 29.56 dB & \footnotesize DCT / 28.27 dB  & \footnotesize DCT/DCT2net / 29.45 dB    \\

 &&&&\\
 \includegraphics[scale=0.26]{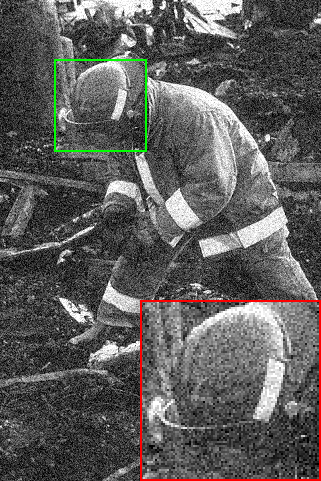} &
\includegraphics[scale=0.26]{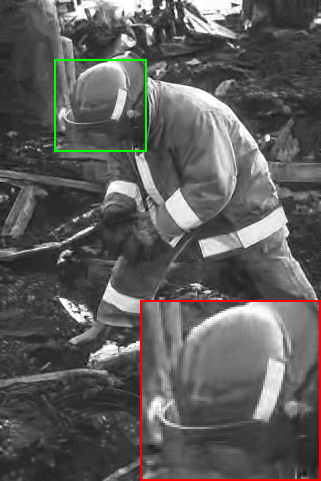} &
\includegraphics[scale=0.26]{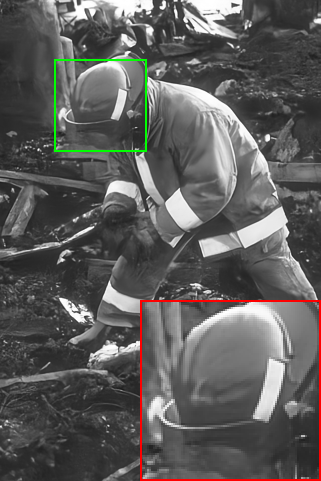} &
\includegraphics[scale=0.26]{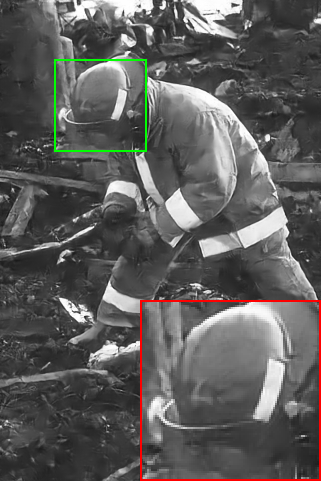} &
\includegraphics[scale=0.26]{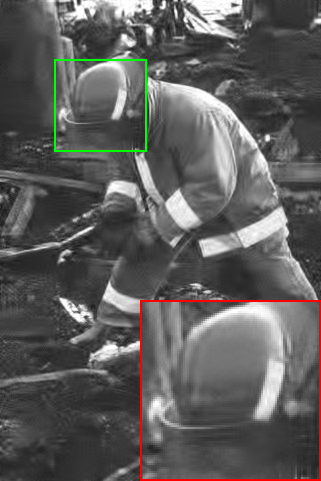} &
\includegraphics[scale=0.26]{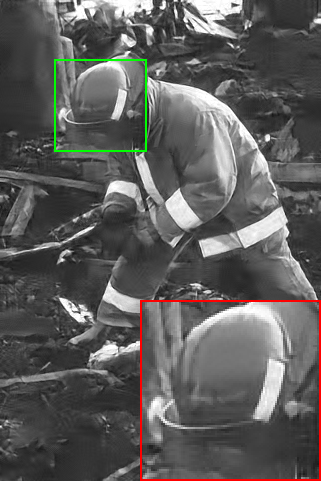} \\
 \footnotesize Noisy / 20.17 dB  &   \footnotesize BM3D / 26.86 dB  &    \footnotesize DnCNN / 27.76 dB  & \footnotesize LKSVD$_{1,8,256}$ / 27.39 dB & \footnotesize DCT / 25.45 dB  & \footnotesize DCT/DCT2net / 26.90 dB     \\

&&&&\\
 \includegraphics[scale=0.26]{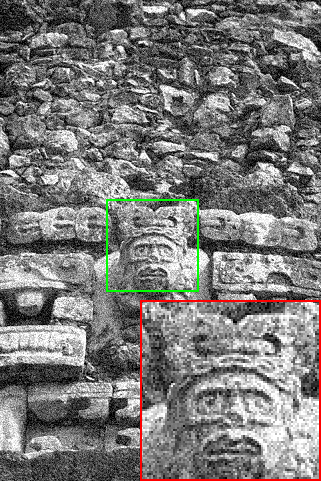} &
\includegraphics[scale=0.26]{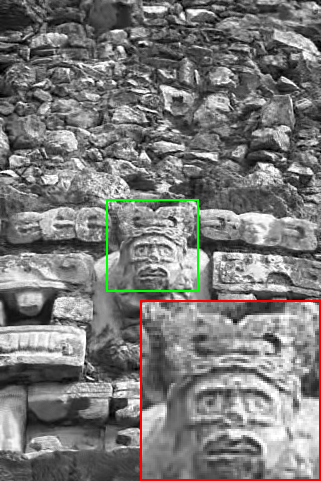} &
\includegraphics[scale=0.26]{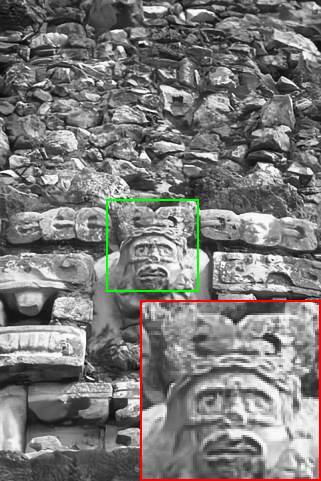} &
\includegraphics[scale=0.26]{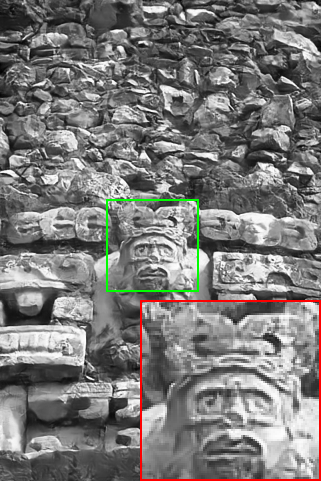} &
\includegraphics[scale=0.26]{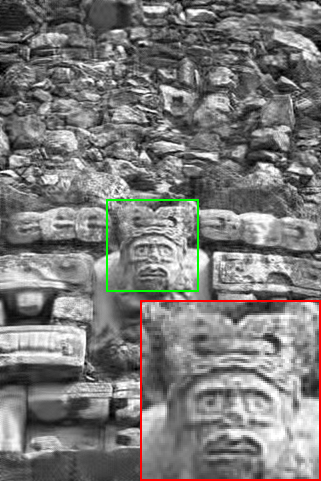} &
\includegraphics[scale=0.26]{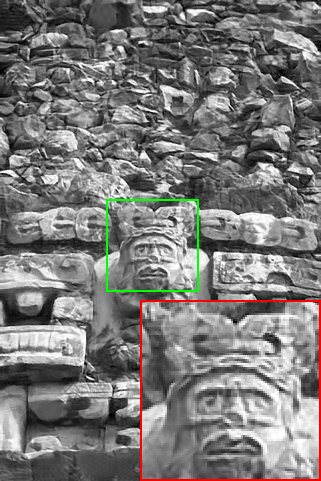} \\
 \footnotesize Noisy / 20.17 dB  &   \footnotesize BM3D / 24.36 dB  &    \footnotesize DnCNN / 24.80 dB  & \footnotesize LKSVD$_{1,8,256}$ / 24.63 dB & \footnotesize DCT / 23.31 dB  & \footnotesize DCT/DCT2net / 24.45 dB      \\

\end{tabular}

\caption{Denoising results (in PSNR) of some images from BSD68 corrupted with white Gaussian noise and $\sigma=20$.}
\label{photo}
\end{figure*}
\addtolength{\tabcolsep}{5pt}

\section{Experiments}
\label{section5}

In this section, we describe the experiments conducted to train our model DCT2net. Moreover, we provide comparisons with traditional and deep-learning-based state-of-the-art algorithms. The code and pre-trained models can be downloaded here:
\href{https://github.com/sherbret/DCT2net/}{https://github.com/sherbret/DCT2net/}.

\begin{table*}[ht]
\centering
  \caption{The PSNR (dB) results of different methods on Set12 corrupted with white Gaussian noise and $\sigma=15$, $25$ and $50$. The best tow results are highlighted in \textcolor{red}{red} and \textcolor{blue}{blue} colors, respectively.}\label{set12tab}
  
  \begin{tabular}{*{14}{|c}|}
  \hline
 \rowcolor{lightgray} \textbf{Images} & \textit{C.man} & \textit{House} & \textit{Peppers} & \textit{Starfish}& \textit{Monarch} &\textit{Airplane}&\textit{Parrot}&\textit{Lena}&\textit{Barbara}&\textit{Boat}&\textit{Man}&\textit{Couple}&\textit{Average}\\\hline\hline

 Noise Level & \multicolumn{13}{c|}{$\sigma = 15$} \\ \hline
 BM3D \cite{BM3D} & 31.91&\textcolor{blue}{34.93}&32.69&31.14&31.85&31.07&31.37&\textcolor{blue}{34.26}&\textcolor{red}{33.10}&\textcolor{blue}{32.13}&31.92&\textcolor{blue}{32.10}&\textcolor{blue}{32.37}\\\hline
 PEWA \cite{PEWA} &31.88&34.72&32.64&30.85&31.83&31.04&31.29&34.09&\textcolor{blue}{32.73}&31.90&31.81&31.88&32.22 \\\hline

 DnCNN \cite{dncnn} &\textcolor{red}{32.61}&\textcolor{red}{34.97}&\textcolor{red}{33.30}&\textcolor{red}{32.20}&\textcolor{red}{33.09}&\textcolor{red}{31.70}&\textcolor{red}{31.83}&\textcolor{red}{34.62}&32.64&\textcolor{red}{32.42}&\textcolor{red}{32.46}&\textcolor{red}{32.47}&\textcolor{red}{32.86} \\\hline
 
  LKSVD$_{1, 8, 256}$ \cite{deepKSVD} & \textcolor{blue}{32.07} & 34.26 & \textcolor{blue}{32.79} & \textcolor{blue}{31.62} & \textcolor{blue}{32.49} & \textcolor{blue}{31.37} & \textcolor{blue}{31.62} & 34.03 & 31.84 & 31.97 & \textcolor{blue}{32.07} & 31.85 & 32.33\\\hline
 
  DCT \cite{DCT} &30.77&33.56&31.65&30.09&30.62&30.17&30.64&33.44&31.63&31.36&31.04&31.20&31.35\\\hline
 DCT2net   & 31.60 & 34.31 & 32.57 & 31.04 & 31.69 & 30.99 & 31.36 & 33.96 & 31.81 & 31.95 & 31.97 & 31.89 & 32.10 \\\hline
 DCT/DCT2net  & 31.49 & 34.30 & 32.52 & 30.88 & 31.60 & 30.93 & 31.27 & 33.93 & 31.90 & 31.82 & 31.78 & 31.78 & 32.02 \\\hline

 Noise Level & \multicolumn{13}{c|}{$\sigma = 25$} \\ \hline
 BM3D \cite{BM3D} & 29.45 & \textcolor{blue}{32.85} & 30.16 & 28.56 & 29.25 & 28.42 & 28.93  & \textcolor{blue}{32.07} & \textcolor{red}{30.71} & \textcolor{blue}{29.90} & 29.61 & \textcolor{blue}{29.71} & \textcolor{blue}{29.97} \\\hline
 PEWA \cite{PEWA} & 29.48 & 32.77 &\textcolor{blue}{30.30}&28.13&29.13&28.41&28.90&31.89&\textcolor{blue}{30.28}&29.65&29.50&29.48&29.83\\\hline

 DnCNN \cite{dncnn} &\textcolor{red}{30.18}&\textcolor{red}{33.06}&\textcolor{red}{30.87}&\textcolor{red}{29.41}&\textcolor{red}{30.28}&\textcolor{red}{29.13}&\textcolor{red}{29.43}&\textcolor{red}{32.44}&30.00&\textcolor{red}{30.21}&\textcolor{red}{30.10}&\textcolor{red}{30.12}&\textcolor{red}{30.43} \\\hline
 
 LKSVD$_{1, 8, 256}$ \cite{deepKSVD} & \textcolor{blue}{29.49} & 31.99 & 30.19 & \textcolor{blue}{28.76} & \textcolor{blue}{29.73} & \textcolor{blue}{28.75} & \textcolor{blue}{29.09} & 31.67 & 28.86 & 29.66 & \textcolor{blue}{29.65} & 29.33 & 29.76\\\hline

  DCT \cite{DCT} & 28.09 & 31.18 & 29.02 & 27.30 & 27.71 & 27.50 & 28.10 & 31.05 & 28.69 & 28.94 & 28.72 & 28.70 & 28.75 \\\hline

  DCT2net  & 29.29 & 32.20 & 30.15 & 28.45 & 29.16 & 28.48 & 28.96 & 31.76 & 29.16 & 29.71 & 29.64 & 29.51 & 29.71\\\hline

 DCT/DCT2net  & 29.16 & 32.26 & 30.08 & 28.32 & 29.08 & 28.42 & 28.88 & 31.75 & 29.29 & 29.56 & 29.41 & 29.41 & 29.64\\\hline

 Noise Level & \multicolumn{13}{c|}{$\sigma = 50$} \\ \hline
 BM3D \cite{BM3D} & 26.13&\textcolor{blue}{29.69}&26.68&25.04&25.82&25.10&25.90&\textcolor{blue}{29.05}&\textcolor{red}{27.22}&\textcolor{blue}{26.78}&\textcolor{blue}{26.81}&\textcolor{blue}{26.46}&\textcolor{blue}{26.72} \\\hline
 PEWA \cite{PEWA} & 26.25 & 29.29 & \textcolor{blue}{26.69} & 24.53 & 25.46 & 25.07 & 25.82 & 28.83 & \textcolor{blue}{26.58} & 26.64 & 26.67 & 26.02 & 26.49 \\\hline
 DnCNN \cite{dncnn} &\textcolor{red}{27.03}&\textcolor{red}{30.00}&\textcolor{red}{27.32}&\textcolor{red}{25.70}&\textcolor{red}{26.78}&\textcolor{red}{25.87}&\textcolor{red}{26.48}&\textcolor{red}{29.39}&26.22&\textcolor{red}{27.20}&\textcolor{red}{27.24}&\textcolor{red}{26.90}&\textcolor{red}{27.18} \\\hline
 LKSVD$_{1, 8, 256}$ \cite{deepKSVD} & \textcolor{blue}{26.26} & 28.53 & 26.52 & \textcolor{blue}{25.12} & \textcolor{blue}{26.00} & \textcolor{blue}{25.31} & \textcolor{blue}{25.93} & 28.32 & 24.75 & 26.55 & 26.68 & 26.07 & 26.34\\\hline
 
  DCT \cite{DCT} &  24.67 & 27.73 & 25.48 & 23.93 & 24.10 & 24.05 & 24.78 & 27.71 &  24.98 & 25.81 &  26.01 & 25.55 & 25.40 \\\hline

 DCT2net   & 26.20 & 28.78 & 26.59 & 24.86 & 25.54 & 25.15 & 25.91 & 28.55 & 25.53 & 26.62 & 26.70 & 26.27 & 26.39 \\\hline

 DCT/DCT2net & 26.20 & 29.05 & 26.48 & 24.74 & 25.41 & 25.15 & 25.89 & 28.63 & 25.73 & 26.47 & 26.56 & 26.20 & 26.38 \\\hline

\end{tabular}
\end{table*}

\begin{table*}[ht]
\centering
  \caption{The average PSNR (dB) results of different methods on BSD68 corrupted with white Gaussian noise and $\sigma=15$, $25$ and $50$. The best two results are highlighted in \textcolor{red}{red} and \textcolor{blue}{blue} colors, respectively.}\label{bsd68tab}

  \begin{tabular}{*{8}{|c}|}
  \hline
  \rowcolor{lightgray}  \textbf{Methods} & BM3D & PEWA &  DnCNN   & LKSVD$_{1, 8, 256}$  & DCT &  DCT2net & DCT/DCT2net \\\hline\hline
 
  {\tiny $\sigma = 15$} & 31.07 & 31.04 & \textcolor{red}{31.72} & \textcolor{blue}{31.33} & 30.32 &31.09& 30.97\\\hline
 {\tiny $\sigma = 25$} & 28.57 & 28.52  & \textcolor{red}{29.23} & \textcolor{blue}{28.76} & 27.76 &28.64& 28.53 \\\hline
 {\tiny $\sigma = 50$} & 25.62 & 25.53  & \textcolor{red}{26.23} & \textcolor{blue}{25.68} & 24.86 &\textcolor{blue}{25.68}& 25.59\\\hline
\end{tabular}
\end{table*}

\subsection{Training Settings}

We trained our DCT2net on 400 gray-scale images from the Berkeley segmentation dataset (BSDS) \cite{berkeley} where synthetic Gaussian noise with zero mean and a random standard deviation $\sigma$ taken in  $\in [1, 55]$ was added in order to create our pairs $(\boldsymbol{x}_i, \boldsymbol{y}_i)_{i \in \{ 1, \ldots, N\}}$ with $N \gg 400$ (the $\boldsymbol{x}_i$ are redundant). Contrary to numerous deep learning models, our network can adapt to the level of noise as the differentiable hard shrinkage function depends on $\sigma$, so that we train our model only once for all levels of noise at the same time.

During training, we randomly  sample cropped images from the training set of size $128 \times 128$ with a mini-batch size of $32$. We use horizontal and vertical flipping as well as random rotations $\in \{ 0^{\circ},90^{\circ},180^{\circ},270^{\circ}\}$ as further data augmentation. In total, $400 \times 665$ overlapping patches from $400$ clean images are used for training. The mean squared error was used as loss function  and we used Adam optimizer \cite{adam}. The learning rate was set to $10^{-3}$ and decreased exponentially to $10^{-5}$  during the $15$ epochs required for convergence. Note that we initialized the weights of our networks according to the original discrete cosine transform given by (\ref{dctcoeff}). Initializing the weights randomly, for example with a Xavier initialization as it is usually done with deep neural networks, only slows down the time for convergence. As for the parameter $m$ specifying the degree of approximation of the hard shrinkage function $\varphi_{\lambda}$, we took $m=32$. Training  a  model  took  approximately 8 hours  with  a  GeForce RTX 2080 Ti.

Note that a small improvement of our hybrid solution DCT/DCT2net can be obtained by training DCT2net only on parts of images where the learned transform will be applied. In practice, this is done by pre-computing binary masks $b_i$ according to the classification proposed for every image of the external dataset and solving:

\begin{equation}
   \boldsymbol{P}^{*} = \arg \min_{\boldsymbol{P}} \sum_{i=1}^{N} \| b_i \odot (F_{\boldsymbol{P}}(\boldsymbol{y}_i, \sigma_i) - \boldsymbol{x}_i) \|^2_2 
\end{equation}

\noindent where $F_{\boldsymbol{P}}$ designates the network and $\odot$ is the Hadamard product.

We recall the parameters chosen for the Canny edge detector: lower and upper thresholds are set once and for all respectively to $0.1$ and $0.2$ and a supplementary dilation operation is performed using a kernel of size $5 \times 5$.

\begin{table*}[ht]
\centering
  \caption{Running time (in seconds) of different methods for denoising images with size 256$\times$256, 512$\times$512 and 1,024$\times$1,024. Run times are given on CPU (left) and GPU (right) when possible.}\label{time}

  \begin{tabular}{*{8}{|c}|}
  \hline
  \rowcolor{lightgray}  \textbf{Image size} & BM3D\cite{BM3D} & PEWA\cite{PEWA} &  DnCNN\cite{dncnn} &  LKSVD$_{1, 8, 256}$ \cite{deepKSVD} &  DCT 16$\times$16\cite{DCT} &   DCT2net  & DCT/DCT2net \\\hline\hline
 
 256$\times$256 & 1.73 & 38.85 & 0.87\;/\;0.010 & 1.15\;/\;0.020 & 0.49\;/\;0.005 & 0.39\;/\;0.005  & 1.05 \\\hline
 512$\times$512 & 6.65 & 190.82 & 3.47\;/\;0.037 & 5.78\;/\;0.082 & 2.02\;/\;0.037 & 1.56\;/\;0.027  & 4.08  \\\hline
 1,024$\times$1,024 & 26.90 & 803.76 & 18.35\;/\;0.145 & 25.78\;/\;0.332 & 8.70\;/\;0.161 & 5.88\;/\;0.112 & 16.87  \\\hline
 
\end{tabular}
\end{table*}

\begin{table}[t]
\centering
  \caption{Model  complexities  comparison  of  our  proposed method with two popular networks}\label{parameters}

  \begin{tabular}{*{4}{|c}|}
  \hline
  \rowcolor{lightgray}  \textbf{Methods} & DnCNN &  LKSVD$_{1, 8, 256}$ &  DCT2net  \\\hline\hline
 
Number of layers & 17 & 5 & 2 \\\hline
Number of parameters & 556,096 & 35,138 & 28,561 \\\hline
 
\end{tabular}
\end{table}

\subsection{Results on test datasets}

We tested the denoising performance of our architecture on two well-known datasets: Set 12 and BSD68. According to our experiments, the best model for DCT/DCT2net in terms of performance (i.e., PSNR) and subjective visual quality is obtained with a patch size of $13$. Larger sizes of patch only bring negligible enhancement for a complexity much more important.

Tables \ref{set12tab} and \ref{bsd68tab} compare the performance of traditional and deep-learning-based state-of-the-art algorithms with our model. We  compare  our  DCT2net    with BM3D \cite{BM3D} and PEWA \cite{PEWA}, both state-of-the-art traditional methods that exploit self-similarity and DCT decomposition. We also compare DCT2net to related algorithm Deep K-SVD \cite{deepKSVD}. Scetbon et al. \cite{deepKSVD} proposed multiple models that depend on the patch size, the dictionary  size and the number  of  denoising  steps. In what follows, we consider the smallest model for which an implementation is given by the authors and which is denoted LKSVD$_{1, 8, 256}$. We performed the training by ourselves for each noise level, as no pretrained models were supplied by the authors.

We can notice that DCT2net achieves comparable performances with state-of-the-art traditional algorithms on both Set12 and BSD68 datasets, outperforming its original counterpart DCT\footnote{The non-adaptive version of DCT denoiser was considered as it produced a slightly higher PSNR, despite its poor subjective visual quality.} \cite{DCT}. Unlike BM3D and PEWA and other algorithms such as NL-Bayes \cite{nlbayes}, DCT2net is a very simple one-pass algorithm, able to produce similar performances to Deep K-SVD for high noise level while it is not trained specifically to address such challenging situations.

Beyond the performance assessed with the PSNR criterion, nothing can replace the subjective assessment of a human eye. On this criterion, our DCT/DCT2net can hold its own against established methods such as BM3D as shown in Figure \ref{photo}.
The use of the traditional DCT on flat areas produces, for example, a better-looking sky than Deep K-SVD or BM3D when applied to the \textit{Castle} image. Unsurprisingly, DCT/DCT2net (based on two layers) cannot compete with very deep neural networks such that DnCNN \cite{dncnn} but it is faster (see below and Table \ref{time}).

Nevertheless, the performance has to be put in perspective with the complexity of the model which is studied in the next subsection.

\subsection{Complexity and low-cost training}

We want to emphasize that DCT2net is very light and fast compared to its traditional and deep-learning-based counterparts. In Table \ref{parameters}, we reported the complexity in terms of layers numbers and parameters. The number of parameters of DCT2net represents only 5\% of the total of parameters of DnCNN. Moreover, the underlying parameters are the same whatever the noise level is, which is not the case for DnCNN or Deep K-SVD where the  models have to be trained from scratch for every noise level\footnote{Although solutions for handling multiple noise levels within the same network were proposed, including a noisemap at the network entry by the authors of \cite{ffdnet}.}

In addition to the number of parameters, the executing time is a crucial feature of denoising algorithms. Table \ref{time} is provided for information purposes only, as the implementation, the language used and the machine on which the code is run, highly influence the execution time. The CPU used is a 2,3 GHz Intel Core i7 and the GPU is a GeForce RTX 2080 Ti. We used the IPOL implementation \cite{ipol} for BM3D \cite{BM3D} and the implementation provided by the authors for the others algorithms, except for DCT \cite{DCT} that we re-implemented with Pytorch on our own, leveraging the network unfolding scheme already used in \cite{deepKSVD}. By the way,  DCT2net can be easily adapted to this specific unfolding implementation, as there is no difference between DCT2net and DCT denoiser, apart from the underlying bases. 

We also tried to train our network on fewer images than the original BSD400 dataset. As our network relies on a two-layers architecture, it is less prone to overfitting and the learning can be performed only from several dozens of images. With 10 and 40 images, corresponding to $10 \times 665$ and $40 \times 665$  overlapping patches  of size $128\times128$ respectively, our DCT/DCT2net achieves almost the same performance with no visual difference. By way of comparison, training a model with 40 images takes less than one hour with a GeForce RTX 2080 Ti and the average PSNR values on Set12 are 32.07dB, 29.69dB and 26.39dB, for $\sigma=15$, $25$ and $50$ respectively.

\section{Discussion and conclusion}

DCT2net is one of the first attempts to create an interpretable shallow CNN for image denoising. Sure enough, the  performance  of  DCT2net still falls short when compared to state-of-the-art deep-learning-based methods such as DnCNN \cite{dncnn}. However, manipulating shallow networks has much to offer. Beyond the fact that they are extremely fast as the number of hidden layers is limited, these networks challenge us to think differently our approach to neural networks and encourage us to be more creative than the traditional "Transform-BatchNorm-ReLU" repeated dozens of times. With shallow networks, the activation function must be carefully designed to best match its purpose. Thus, during the training phase, DCT2net uses an approximation of a hard shrinkage function as activation function that depends on the noise level. This is, to the best of our knowledge, the first time such a function is used in a CNN.

Moreover, DCT2net is fully interpretable, unlike Deep K-SVD \cite{deepKSVD} that uses a multilayer perceptron (MLP) ahead of its sparsity-based network. This interpretability is an important advantage, making the method more robust. At the end of the optimization process, it is possible to check what the network has just learned and, in the case of DCT2net, to directly display the learned basis. By easily exploring the different steps of the process, some usages, usually taken for granted, are disproved by the machine. Thus, we were surprised to realize that the aggregation step that is common in denoising methods based on patches, is not a basic post-processing step but can be fully integrated in the denoising process to considerably improve the performance.

This study shows that signal processing methods such as the popular DCT denoising algorithm can have a comeback by improving the transform involved through deep learning framework. We showed that fully interpretable CNNs can be designed, for which denoising performances compare favorably with state-of-the-art traditional algorithms. We hope that our work will open the door to new architectures, more reliable and understandable for the human brain.

\appendices
\section{Why is taking multiple thresholds useless ?}
\label{appendix2}

In the definition of DCT2net (and traditional DCT), a unique threshold $\lambda$, dependent on the level of noise $\sigma$, is applied to all the coefficients of the vector $\boldsymbol{P}^{-1} \boldsymbol{y}$, corresponding to the frequency representation of the signal $\boldsymbol{y}$. One may wonder what would bring a different threshold for every coefficient, replacing the function $\varphi_{\lambda}$ by $\varphi_{\lambda_1, \ldots, \lambda_n}$ defined by:
$$\forall \boldsymbol{x} \in \mathbb{R}^n, \; \varphi_{\lambda_1, \ldots, \lambda_n}( \boldsymbol{x}) = (\varphi_{\lambda_1}(x_1), \ldots, \varphi_{\lambda_n}(x_n))$$
As a matter of fact, defining multiple thresholds is useless as the matrix $\boldsymbol{P}$ and the threshold values $\lambda_1, \ldots, \lambda_n$ can be "encoded" in a single matrix as explained by the following result.
\begin{proposition}
Let $\lambda_1, \ldots, \lambda_n > 0$ be $n$ values of threshold and  $\boldsymbol{\Lambda}= \text{diag}(\lambda_1, \ldots, \lambda_n)$.

 $\forall \boldsymbol{P} \in \mathcal{GL}_n(\mathbb{R}), \forall \boldsymbol{y} \in \mathbb{R}^n, \forall \sigma > 0,$ 
$$\boldsymbol{P} \varphi_{\lambda_1 \sigma, \ldots, \lambda_n \sigma}(\boldsymbol{P}^{-1} \boldsymbol{y}) = (\boldsymbol{P \Lambda}) \varphi_{\sigma}((\boldsymbol{P \Lambda})^{-1} \boldsymbol{y})$$

\label{prop1}
\end{proposition}

\begin{proof}
The result can be easily derived thanks to the property on hard shrinkage functions, stating that for two levels of threshold $\lambda$ and $\lambda'$, we have $\varphi_{\lambda}(x) = \frac{\lambda}{\lambda'} \varphi_{\lambda'}( \frac{\lambda'}{\lambda}x)$.
\end{proof}

\section{Direct technique to derive an orthonormal matrix for DCT2net}
\label{appendix1}

In addition to the technique relying on the introduction of a regularization term, we expose here a direct technique that is based on the following lemma.

\begin{lemma}
Let $\mathcal{O}_n(\mathbb{R})$ be the set of orthonormal matrices, $\mathcal{GL}_n(\mathbb{R})$ the set of invertible matrices and $\mathcal{S}_n^{++}$ the set of symmetric positive definite matrices of size $n \times n$. Then,
$$ \mathcal{O}_n(\mathbb{R}) = \left\{\boldsymbol{M} \left(\sqrt{\boldsymbol{M}^T \boldsymbol{M}}\right)^{-1} \; | \; \boldsymbol{M} \in \mathcal{GL}_n(\mathbb{R}) \right\}.$$

\noindent where $\sqrt{\boldsymbol{A}}$ designates the only matrix of $\mathcal{S}_n^{++}$ such that $\boldsymbol{A} = \sqrt{\boldsymbol{A}} \times \sqrt{\boldsymbol{A}}$ (exists and is unique if $\boldsymbol{A} \in \mathcal{S}_n^{++}$). 
\label{lemme1}
\end{lemma}

\begin{proof} 
First of all, $\forall \boldsymbol{M} \in \mathcal{GL}_n(\mathbb{R})$, $\boldsymbol{M}^T \boldsymbol{M}  \in \mathcal{S}_n^{++}$. Moreover, $\forall \boldsymbol{A} \in \mathcal{S}_n^{++}$, $\boldsymbol{A}$ is invertible (with  $\boldsymbol{A}^{-1}  \in \mathcal{S}_n^{++}$). Therefore, for all $\boldsymbol{M} \in \mathcal{GL}_n(\mathbb{R})$, $\boldsymbol{M} (\sqrt{\boldsymbol{M}^T \boldsymbol{M}})^{-1}$ is well defined.

Now, by double inclusion:

\noindent $(\subset)$: Let $\boldsymbol{Q} \in \mathcal{O}_n(\mathbb{R})$. We set $\boldsymbol{M} = \boldsymbol{Q} \in \mathcal{GL}_n(\mathbb{R})$. 

\noindent $\sqrt{\boldsymbol{M}^T \boldsymbol{M}} = \boldsymbol{I}$ is invertible and $\boldsymbol{Q} = \boldsymbol{M} (\sqrt{\boldsymbol{M}^T \boldsymbol{M}})^{-1}$.

\noindent $(\supset)$: Let $\boldsymbol{M} \in \mathcal{GL}_n(\mathbb{R})$ and $\boldsymbol{Q} = \boldsymbol{M} (\sqrt{\boldsymbol{M}^T \boldsymbol{M}})^{-1}$. Using that  for all $\boldsymbol{A} \in \mathcal{S}_n^{++}, (\sqrt{\boldsymbol{A}})^{-1} = \sqrt{\boldsymbol{A}^{-1}}$, we have:
\begin{tabular}{rcl}
$\boldsymbol{Q} \boldsymbol{Q}^T$ &$=$& $\boldsymbol{M} (\sqrt{\boldsymbol{M}^T \boldsymbol{M}})^{-1}  (\sqrt{\boldsymbol{M}^T \boldsymbol{M}})^{-1}  \boldsymbol{M}^T$ \\
&$=$& $\boldsymbol{M} \sqrt{(\boldsymbol{M}^T \boldsymbol{M})^{-1}}  \sqrt{(\boldsymbol{M}^T \boldsymbol{M})^{-1}}  \boldsymbol{M}^T$ \\
&$=$& $\boldsymbol{M} (\boldsymbol{M}^T \boldsymbol{M})^{-1}  \boldsymbol{M}^T$ \\
&$=$& $\boldsymbol{M} \boldsymbol{M}^{-1}  (\boldsymbol{M}^T)^{-1}  \boldsymbol{M}^T$ \\
&$=$& $\boldsymbol{I}$
\end{tabular}

\noindent hence, $\boldsymbol{Q} \in \mathcal{O}_n(\mathbb{R})$.
\end{proof}

Let $F_{\boldsymbol{P}}$ denote the network DCT2net where $\boldsymbol{P}$ is the learned transform. The direct technique consists in solving the following optimization problem:
\begin{equation}
      \boldsymbol{M}^* = \arg \min_{\boldsymbol{M} \in \mathcal{GL}_{p^2}(\mathbb{R})} \sum_{i=1}^{N} \|  F_{\boldsymbol{M} \left(\sqrt{\boldsymbol{M}^T \boldsymbol{M}}\right)^{-1}}(\mathbf{y}_i, \sigma_i) - \mathbf{x}_i \|^2_2 
\end{equation}
 
\noindent Similarly to the unconstrained formulation of DCT2net (\ref{equation4}), the optimization problem is solved by stochastic gradient descent, leveraging the power of automatic differentiation in modern machine learning libraries such as Pytorch. The learned transform $\boldsymbol{P}^{*}$ is reconstructed at the end and is guaranteed to be orthonormal thanks to Lemma \ref{lemme1}:
$$\boldsymbol{P}^{*} =\boldsymbol{M^*} \left(\sqrt{\boldsymbol{M^*}^T \boldsymbol{M^*}}\right)^{-1}$$

\section{Link between orthonormal matrices and orthogonal ones in DCT2net}
\label{appendix3}

Although often used as synonyms in the literature, a clear distinction between orthonormal matrices and orthogonal ones is made in this paper.

\begin{definition}
Let $\boldsymbol{P}$ be a matrix of size $n \times n$. 
\begin{itemize}
    \item $\boldsymbol{P}$ is an orthonormal matrix, and we note $\boldsymbol{P} \in \mathcal{O}_n(\mathbb{R})$, if $\boldsymbol{P}^T \boldsymbol{P} = \boldsymbol{P} \boldsymbol{P}^T = \boldsymbol{I}_n$.
     \item $\boldsymbol{P}$ is an orthogonal matrix, and we note $\boldsymbol{P} \in \mathcal{O}^g_n(\mathbb{R})$, if $\boldsymbol{P}^T \boldsymbol{P}= \boldsymbol{D}$, with $\boldsymbol{D}$ an invertible diagonal matrix.
\end{itemize}
\end{definition}

In other words, a matrix $\boldsymbol{P}$ is said to be \textit{orthonormal} if its columns $c_1, \ldots, c_n$ have the property: $\forall i, j \in \{1, \ldots ,n\}, \; \langle c_i, c_j \rangle = \delta_{i,j}$ where $\delta_{i,j}$ is the Kronecker delta. The \textit{orthogonality} property is less restrictive as its columns must satisfy $\forall i, j \in \{1, \ldots ,n\}, \;  \langle c_i, c_j \rangle = 0 \Leftrightarrow i \neq j $.

\bigskip

Taking $\boldsymbol{P} \in \mathcal{O}_n(\mathbb{R})$ with multiple values of threshold amounts to considering only one value of threshold with $\boldsymbol{P} \in \mathcal{O}^g_n(\mathbb{R})$ and conversely. Indeed, let  $\boldsymbol{P} \in \mathcal{O}^g_n(\mathbb{R})$. There exists $\boldsymbol{D}$ an invertible diagonal matrix such that $\boldsymbol{P}^T  \boldsymbol{P}  = \boldsymbol{D}$. We can write $\boldsymbol{P} = \boldsymbol{Q} \sqrt{\boldsymbol{D}}$ with $\boldsymbol{Q} = \boldsymbol{P} (\sqrt{\boldsymbol{D}})^{-1} \in \mathcal{O}_n(\mathbb{R})$.  Now applying Prop. \ref{prop1} for $\lambda_i = \sqrt{\boldsymbol{D}}_{i, i} > 0$ and $\boldsymbol{Q}$ gives that
$\forall \boldsymbol{y} \in \mathbb{R}^n, \forall \sigma > 0,$ 
$$\boldsymbol{P} \varphi_{\sigma}(\boldsymbol{P }^{-1} \boldsymbol{y}) =  \boldsymbol{Q} \varphi_{\lambda_1 \sigma, \ldots, \lambda_n \sigma}(\boldsymbol{Q}^{-1} \boldsymbol{y})$$

\section*{Acknowledgment}

This work was supported by Bpifrance agency (funding) through the LiChIE contract. Computations  were performed on the Inria Rennes computing grid facilities partly funded by France-BioImaging infrastructure (French National Research Agency - ANR-10-INBS-04-07, “Investments for the future”).

We would like to thank R. Fraisse (Airbus) for fruitful  discussions.

\end{document}